\documentclass[]{aa}  

\usepackage{graphicx}
\usepackage{txfonts}
\newcommand{\PFAA}{\texttt{RunC}} 

\newcommand{\Jr}{$J_\mathrm{R}$}
\newcommand{\oJr}{$\overline{J_\mathrm{R}}$}
\newcommand{\mathoJr}{\overline{J_\mathrm{R}}}
\newcommand{\mathJr}{J_\mathrm{R}}

\newcommand{\Er}{$E_\mathrm{R}$}
\newcommand{\mathEr}{E_\mathrm{R}}

\newcommand{\Lz}{$L_\mathrm{z}$}
\newcommand{\oLz}{$\overline{L_\mathrm{z}}$}
\newcommand{\mathoLz}{\overline{L_\mathrm{z}}}
\newcommand{\mathLz}{L_\mathrm{z}}

\newcommand{\kms}{km\,s$^{-1}$}
\newcommand{\Lzunit}{kpc\,km\,s$^{-1}$}
\newcommand{\msol}{M$_\sun$}
\newcommand{\mtot}{M$_\mathrm{tot}$}
\newcommand{\omegaunit}{km$\,$s$^{-1}$\,kpc$^{-1}$}

\newcommand{\RA}{\texttt{Subset~A}}
\newcommand{\RB}{\texttt{Subset~B}}

\newcommand{\UHRb}{UHR$_\mathrm{B}$}
\newcommand{\CRb}{CR$_\mathrm{B}$}
\newcommand{\OLRb}{OLR$_\mathrm{B}$}
\newcommand{\ILRi}{ILR$_\mathrm{iS}$}

\newcommand{\CRi}{CR$_\mathrm{iS}$}
\newcommand{\OLRi}{OLR$_\mathrm{iS}$}
\newcommand{\ILRs}{ILR$_\mathrm{oW}$}
\newcommand{\UHRs}{UHR$_\mathrm{oW}$}
\newcommand{\CRs}{CR$_\mathrm{oW}$}

\begin{document}

   \title{Diffusion of radial action in a galactic disc}

   \author{Herv\'e Wozniak \inst{1} }

   \institute{LUPM, Univ Montpellier, CNRS, Montpellier,
     France\\ \email{herve.wozniak@umontpellier.fr} }

   \date{Received 17 July 2020 / Accepted 1 September 2020}

 
  \abstract{Stellar migration of the galactic disc stars has been
    invoked to explain the dispersion of stellar metallicity observed
    in the solar neighborhood.}  { We seek to identify the dynamical
    mechanisms underlying stellar migration in an isolated galaxy disc
    under the influence of a bar. Our approach is to analyze the
    diffusion of dynamical quantities.} {We extend our previous work
    by exploring Chirikov's diffusion rate (and derived timescale) of
    the radial action \Jr\ in an idealised N$-$body simulation of an
    isolated disc galaxy. We limit our study to the evolution of the
    disc region well after the formation of the bar, in a regime of
    adiabatic evolution.}  {The \Jr\ diffusion timescales
    $T_\mathrm{D}(\mathJr)$ is less than 3~Gyr for roughly half the
    galaxy mass. It is always much shorter than the angular momentum
    diffusion timescale $T_\mathrm{D}(\mathLz)$ outside the stellar
    bar. In the disc, $\langle T_\mathrm{D}(\mathJr)\rangle \sim
    1$~Gyr.  All non-axisymmetric morphological structures
    characteristic of resonances and waves in the disc are associated
    to particles with $T_\mathrm{D}(\mathJr) < 3$~Gyr and
    $T_\mathrm{D}(\mathLz) > 10$~Gyr. Short $T_\mathrm{D}(\mathJr)$
    can be explained by the gradual decircularisation of initially
    circular orbits ($\mathJr=0$) under the effect of intermittent ILR
    scattering by wave trains propagating in the disc, well beyond the
    bar OLR. This leads to a moderate secular heating of the disc
    beyond the bar OLR for 7~Gyr, comparable to solar neighbourhood
    observations. The complex multi-wave structure, mixing permanent
    and intermittent modes, allows for multiple resonance overlaps.}{}


   \maketitle
%
\section{Introduction}
\label{sec:introduction}

In \citet{2020ApJ...889...81W} we used for the first time the
formulation of the diffusion rates introduced by
\citet{1979PhR....52..263C}, applied to both specific energy $E$ and
angular momentum \Lz\ in self-consistent N$-$body experiments of
isolated galactic discs.  Using the same definition, we extend our
previous work by now focusing on the radial action \Jr.

\Jr\ has been introduced in the galactic problem by
\citet{1966MNRAS.133...47F} and \citet{1971ApJ...166..275K}. In the
context of the epicycle approximation, $\mathJr = \mathEr / \kappa$, where
$\kappa$ is the epicycle radial frequency and \Er\ the specific
radial kinetic energy. This approximation is only valid if $\mathJr
\ll \mathLz$ or for near-circular orbits \citep{1971ApJ...166..275K}.

\Jr\ has been considered as a thermometer for measuring the disc
heating in the radial direction. Whether the stellar disc heats or not
during the secular exchange of angular momentum is an open issue. In
particular, it is a question of being able to determine the relative
importance of two phenomena, churning and blurring, for stellar
migration \citep{2009MNRAS.396..203S,2015A&A...578A..58H}. In the
context of epicycle approximation, the effect of blurring is to
increase the amplitude of epicycle motion around a fixed
guiding-centre radius (thus generating no net radial migration in
principle), whereas that of churning is to move the guiding/mean
radius inwards or outwards.

According to \citet{2002MNRAS.336..785S}, any variation of \Jr\ in a
stellar disc excited by a constant non-axisymmetric perturbation of
pattern speed $\Omega_\mathrm{p}$, can be related to $\Delta\mathLz$
as:
\begin{equation}
  \label{eq:djdlde}
  \Delta\mathJr = \frac{\Omega_\mathrm{p}-\Omega}{\kappa}\Delta\mathLz
\end{equation}
where $\Omega$ is the orbit/particle angular frequency. At corotation
$\Omega = \Omega_\mathrm{p}$ so that there is no \Jr\ variation,
whereas at Lindblad resonances with an $m-$armed perturbation,
\begin{equation}
  \label{eq:djdl}
  \Delta\mathJr /  \Delta\mathLz = \mp 1 / m.
\end{equation}
This applies to any kind of disturbance, such as bar or spiral(s). It
can be generalized to higher order resonances, so that $\Delta\mathJr
/ \Delta\mathLz = \mp l / m$ for any $l\,^{th}$-order resonance, where
$l=0$ is the corotation.

To complete the picture, it is necessary to add a last relation:
\Jr\ is proportional to the square of the epicycle amplitude
\citep{2008gady.book.....B}. Therefore, variations in \Lz\ that do not
produce significant variations in \Jr\ can be interpreted as the mark
of the churning mechanism. Corotation scattering is thus expected to
play a major role for this mechanism
\citep{2002MNRAS.336..785S,2012MNRAS.426.2089R}. On the contrary,
variations in \Jr\ that are not correlated with variations in \Lz\ via
Eq.~(\ref{eq:djdlde}) are typical of blurring.  When both \Lz\ and
\Jr\ vary, the situation is much more complex: churning and blurring
are mixed in time-dependent proportions \citep{2015A&A...578A..58H},
scattering can be non-resonant (e.g. with giant molecular clouds) or
made by Lindblad resonances \citep[e.g. the Inner Lindblad
  Resonance][]{2012ApJ...751...44S}.  Moreover, the bar-spirals
resonance overlap \citep{2010ApJ...722..112M,2011A&A...527A.147M} is
expected to increase radial energy so that stars that were originally
on nearly circular orbits ($\mathJr\approx 0$) can move to more
eccentric orbits.  Additional complexity comes from the presence of a
strong bar. Indeed, the congestion of $n/1$ resonances near the
corotation generates a high degree of stochasticity. The diffusion
capacity of a bar corotation could therefore be more important than
that of a spiral.

The introduction of the Chirikov diffusion rate makes it possible to
quantify the impact of the accumulation over time of small
fluctuations in energy, angular momentum and radial action.  This
quantity takes all variations weighted by all
timescales into account. In addition to other methods, it helps to identify the
predominant dynamical mechanism(s), through the associated
timescales. On the other hand, it does not contain any spatial
information. But we have shown \citep{2020ApJ...889...81W} how to get
around this obstacle.

The paper is organized as follows. After reintroducing some basic
notations and concepts in galactic dynamics, and detailing how \Jr\ is
estimated in our simulations (Sect.~\ref{sec:jr_diffusion}), we
present the results on the Chirikov diffusion timescale of \Jr\ in
Sect.~\ref{sec:results}. Section~\ref{sec:df} contains a discussion on
the stellar distribution function as a function of \Lz\ and \Jr, and
its time evolution. Sect.~\ref{sec:zeroevol} describes how circular
orbits of the outer disc absorb angular momentum and gain
eccentricity. A wave analysis is reported in Sect.~\ref{sec:wave}. Our
results are discussed in Sect.~\ref{sec:discussion} whereas
Sect.~\ref{sec:conclusions} summarises our conclusions.

\section{Computing \Jr\ diffusion}
\label{sec:jr_diffusion} 

\subsection{Dynamical concepts}
\label{sec:concepts}

The {\em Chirikov diffusion rate} of \Jr\ for individual particles,
can be defined as :
\begin{equation}
\label{eq:chirikov}
  D_n(J_\mathrm{R}) = \overline{(\Delta\overline{J_\mathrm{R}})^2/\Delta t}.
\end{equation}
Although the original definition deals only with $E$
\citep[][eq. 4.6]{1979PhR....52..263C}, we have extended the
definition to compute $D_n(L_\mathrm{z})$ and, in fact, any measurable
quantity evolving during a time-dependent simulation
\citep{2020ApJ...889...81W}. In Eq.~(\ref{eq:chirikov}) for $n=2$,
$\overline{J_\mathrm{R}}$ is the value of radial action averaged over
a period of $\Delta t_2=10^2$ (in time unit of the system). We can
also estimate a diffusion timescale $T_\mathrm{D}$ by renormalizing
$D_2$ by $\overline{J_\mathrm{R}}^2$, where the time-average is now
computed over the longest timescale (e.g. the experiment length):
\begin{equation}
T_\mathrm{D}(J_\mathrm{R}) = \overline{J_\mathrm{R}}^2 / D_2(J_\mathrm{R}).
\end{equation}

Several methods exist to estimate numerically actions \citep[see][for
  a comprehensive
  review]{2016MNRAS.457.2107S}. \citet{2019MNRAS.482.1525V} provides a
state-of-the-art software package for building dynamical models based
on an action-angle description, including position-velocity and
action-angle conversion routines.  However, the accurate and fast
estimation of action-angle variables remains an open problem,
especially for non-axisymmetric dynamic systems in fast rotation, such
as barred galaxies. The approach by \citet[][]{2014MNRAS.441.3284S}
requires sampling all orbits in such a way that all periods of each
orbit can be represented. In the case of large N$-$body simulations,
this objective is still extremely difficult, if not impossible, to
achieve for strictly computational reasons. Therefore, we have decided
to remain in a more qualitative than quantitative approach, estimating
the radial action as if the orbits of the dynamical system could be
described by the epicycle approximation at each stage of the
evolution of a galaxy. This is certainly the least bad solution,
especially as many averages are used in the calculation of the
Chirikov diffusion rate.

For a nearly circular orbit in the symmetry plane of an axisymmetric
potential of the form
$\Phi(R,z)=\Phi_\mathrm{R}(R)+\Phi_\mathrm{z}(z)$, \Jr\ reduces to
\begin{equation}
    \mathJr=\frac{1}{2\pi}\oint\,dR \left[2\left(E-E_\mathrm{z}-\frac{\mathLz^2}{2R^2}-\Phi_\mathrm{R}\right) \right]^{1/2}
\end{equation}
where $E_\mathrm{z}=\frac{1}{2}\dot{z}^2+\Phi_\mathrm{z}$.  In the
limit of very small radial excursions around the guiding centre
($X=(R_\mathrm{max}-R_\mathrm{min}) \rightarrow 0$), \Jr\ reduces even
further to $\mathJr \rightarrow {\mathEr}/{\kappa} = \frac{1}{2}\kappa
X^2$ where $\kappa$, $\mathEr$ and $X$ are respectively the epicycle
radial frequency, the radial kinetic energy, and the epicycle
amplitude \citep{2008gady.book.....B}.  The integral must be taken
around a full radial period, a quantity largely unknown (and not
accessible) in an N$-$body simulation.  Since we need
$\overline{\mathJr}$ in Eq.~\ref{eq:chirikov} rather than the exact
instantaneous value which is difficult to compute, the following
averaging procedure allows us to obtain such an approximate
expression:
\begin{equation}
\label{eq:jrmean}
\overline{\mathJr} \approx \overline{\left(\frac{\mathEr}{\kappa}\right)}.
\end{equation}
It is convenient to time-average here over $\Delta t_2 = 10^2$ time
units (i.e. 105.4~Myr). Other time windows will be used throughout
this article, and will be clarified in due course.  As we use
Eq.~\ref{eq:jrmean} to study the diffusion of \Jr, we never have
access to an \emph{instantaneous} value of the radial action.

No assumption that leads to the approximation $\mathJr =
\mathEr/\kappa$ is exact when a galaxy is barred. Indeed, in the bar,
the gravitational potential is not vertically separable at any point
and $\kappa$ derived from the axisymmetric part of the potential is
not correct since isopotentials are no longer circular. Clearly, the
epicycle approximation is only correct away from the influence of the
bar. Therefore, in order to test the robustness of our conclusions, we
also used the generalized $\kappa$ formulation due to
\citet{1990A&A...230...55P} and tried to estimate $\overline{\mathJr}$
as the ratio of averages $\overline{\mathEr}/\overline{\kappa}$. The
results quantitatively differ but this does not drastically change our
conclusions. We will mention in the following results most strongly
impacted by our choice for estimating $\overline{\mathJr}$.

\subsection{Numerical implementation}
\label{sec:numerics}

In the following, we will use \PFAA\ simulation of
\citet{2020ApJ...889...81W} as a reference run. The results are
qualitatively identical for the other simulations. Initial stellar
populations are set up to reproduce an idealised disc galaxy. Scale
lengths and scale heights have been chosen such as to shape an initial
axisymmetric disc galaxy with a small but significant bulge. The total
mass is \mtot$=2\times10^{11}$~\msol. All $4\times 10^{7}$ particles
have the same individual mass so that plots expressed in relative
particle number or mass fraction are equivalent. The initial disc size
is 40~kpc for a scalelength of $\approx 4$~kpc.

Initial velocity dispersions are anisotropic solutions of Jeans
equations, with $\sigma_\mathrm{R} = \sigma_\mathrm{z}$ and
$\sigma_{\theta}^2 = \sigma_\mathrm{R}^2 \kappa^2 / (4\Omega^2)$,
where $\sigma_\mathrm{R}$, $\sigma_{\theta}$ and $\sigma_z$ are three
components of the velocity dispersion along respectively the radial,
azimuthal and vertical directions.

We only study the phase after the formation of the bar ($t> 3.16$~Gyr,
until simulation end at $t=10.54$~Gyr), in a regime that can be
considered as adiabatic. Therefore, the bar \emph{formation} mechanism
has no direct influence on our results.  This obviously does not mean
that the bar has no influence at all on the rest of the disc.

Unless otherwise stated, we have excluded from the analyses particles
escaping the 3D log$-$polar grid ($R>100$~kpc or
$\left|z\right|>7.8$~kpc) as soon as they were out by even a single
timestep between $t=3.16$ and $10.54$~Gyr. During the N$-$nody
computation, they have been tracked by ballistic approximation until
they re-enter the grid, to ensure the best possible conservation of
momenta. These particles amount to 0.11~\mtot\ at the end of \PFAA.
The other particles are named `never-escaped' in the following.

\section{Diffusion timescale ($T_\mathrm{D}(J_\mathrm{R})$) results}
\label{sec:results}
\begin{figure}[tb!]
  \centering
  \includegraphics[draft=false,keepaspectratio,width=\hsize]{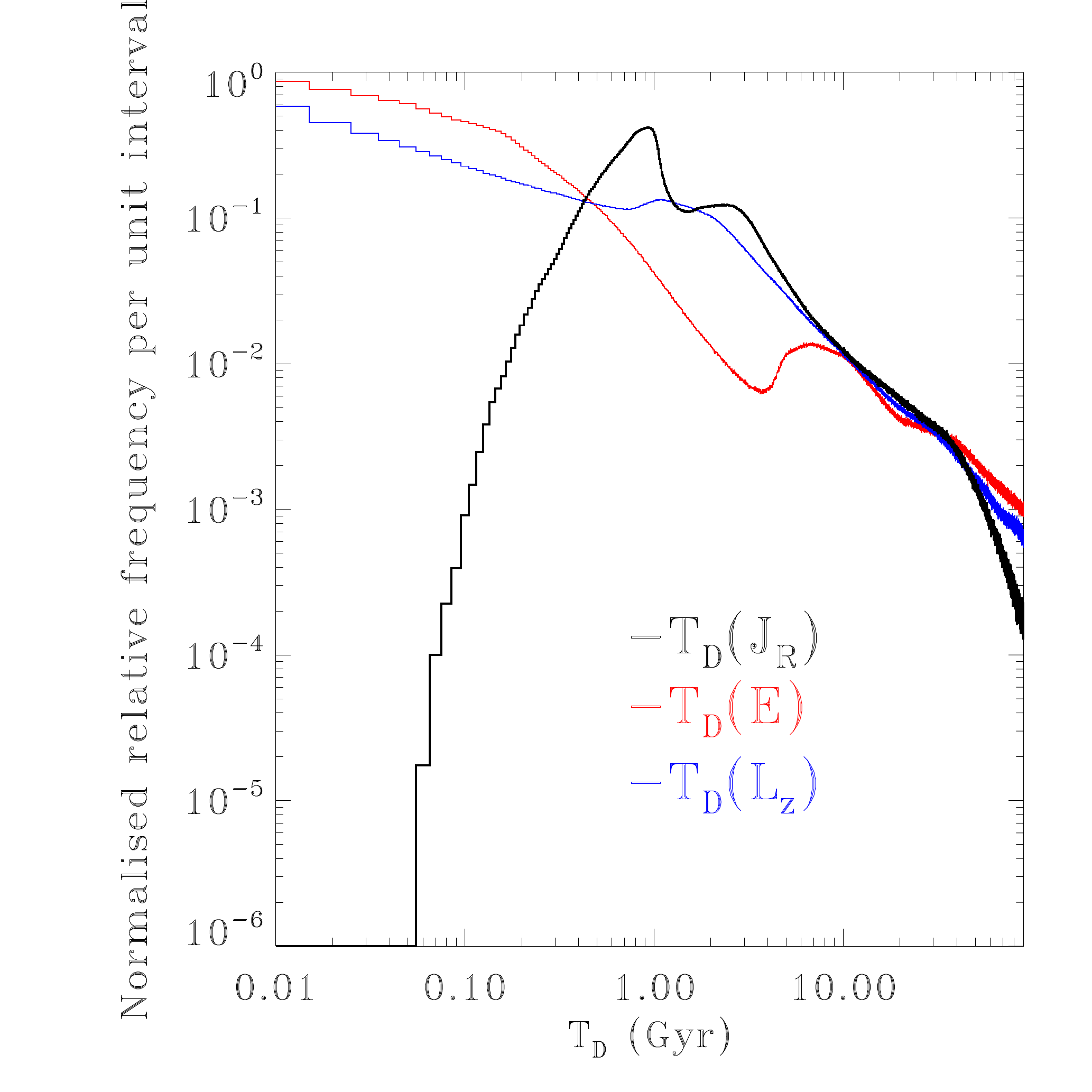}
  \caption{Normalised particle relative frequency per Gyr as a
    function of $T_\mathrm{D}(\mathJr)$ (black histogram). A binsize
    of 0.01~Gyr has been used.  1 particle represents thus a
    probability density of $2.5\times 10^{-6}$~Gyr$^{-1}$.
    Distributions for $T_\mathrm{D}(E)$ (red) and
    $T_\mathrm{D}(\mathLz)$ (blue) are plotted for comparison.}
    \label{fig:timescales}
\end{figure}

The normalised relative frequency distribution (akin to a probability
density function) of $T_\mathrm{D}(J_\mathrm{R})$ is plotted in
Fig.~\ref{fig:timescales}.  For the sake of clarity, we have
restricted this figure to the range $10^{-2} - 90$~Gyr.  Distributions
for $T_\mathrm{D}(\mathJr)$ and $T_\mathrm{D}(E)$ are overplotted for
comparison \citep{2020ApJ...889...81W}.  The distribution shapes are
different, especially for $T_\mathrm{D} \la 0.9$~Gyr.  On the contrary
to $E$ and \Lz, particles with $T_\mathrm{D}(\mathJr) < 100$~Myr are
few. \Jr\ diffusion timescale is never very short.  The maximum of
$T_\mathrm{D}(\mathJr)$ is located at $\approx 0.9$~Gyr, followed by a
plateau up to $\approx 3$~Gyr. $T_\mathrm{D}(\mathJr) = 3$~Gyr is also
the median in mass or number of particles.  On both sides, the
frequency of $T_\mathrm{D}(\mathJr)$ decreases sharply. In other
words, the characteristic diffusion timescale of \Jr\ is remarkably of
the same order of magnitude as dynamical timescales in the galactic
disc. This result contrasts with that obtained for \Lz\ and $E$, for
which $T_\mathrm{D}$ distributions are dominated by short timescales,
outside the local maxima at $T_\mathrm{D}(E)\sim 10$ and
$T_\mathrm{D}(\mathLz)\sim 1$~Gyr. On the other hand, the frequencies
are quite similar beyond 10~Gyr.

\begin{figure}
  \centering
  \includegraphics[draft=false,keepaspectratio,width=\hsize]{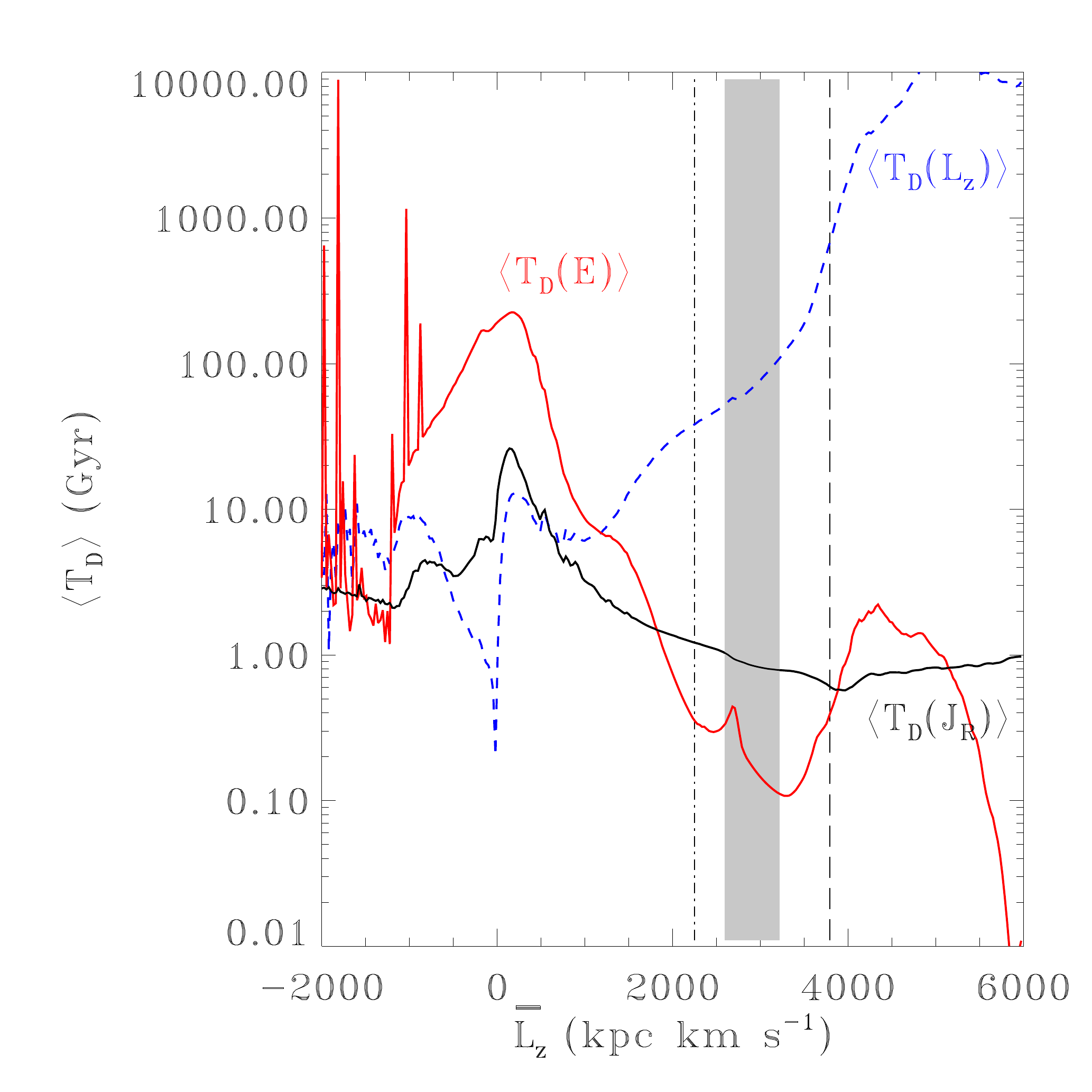}
  \caption{$\langle T_\mathrm{D}(J_\mathrm{R})\rangle$ (black line),
    $\langle T_\mathrm{D}(E)\rangle$ (red line) and $\langle
    T_\mathrm{D}(L_\mathrm{z})\rangle$ (blue line) as a function of
    ${\overline{L_\mathrm{z}}}$. The shaded area delimits the region
    occupied by bar corotation (\CRb) over $\approx 7.4$~Gyr. Vertical
    lines show \Lz\ of circular orbits at the innermost bar UHR
    (\UHRb\ at $t=3.16$~Gyr, dot-dashed) and the outermost bar OLR
    (\OLRb\ at $t=10.54$~Gyr, long-dashed) positions reached.}
  \label{fig:td_lz}
\end{figure}

Fig.~\ref{fig:td_lz} shows $T_\mathrm{D}(\mathJr)$ averaged over sets
of particles (designated by $\langle T_\mathrm{D}(\mathJr)\rangle$ )
sampled by ${\overline{L_\mathrm{z}}}$
ranges. ${\overline{L_\mathrm{z}}}$ is now time-averaged over $\approx
7.4$~Gyr (i.e. from $t=3.16$ to $10.54$~Gyr), and thus must not be
confused with an instantaneous \Lz\ that is not conserved. The range
in $\overline{\mathLz}$ occupied by the bar Lindblad resonances during
the evolution is delimited by the innermost position of the bar UHR
(\UHRb\ at $t=3.16$~Gyr) and the outermost bar OLR (\OLRb\ at
$t=10.54$~Gyr), whereas the range covered by the corotation (\CRb) is
approximately represented by the shaded area. Again, for the sake of
comparison, $\langle T_\mathrm{D}(E)\rangle$ and $\langle
T_\mathrm{D}(L_\mathrm{z})\rangle$ obtained in
\citet{2020ApJ...889...81W} are overplotted.

As shown in \citet{2020ApJ...889...81W}, $\langle
T_\mathrm{D}(E)\rangle$ decreases overall from the centre
($\mathoLz\approx 0$) to the outermost regions, with a strong
depression in the area occupied by the bar Lindblad resonances, while
$\langle T_\mathrm{D}(\mathLz)\rangle$ increases monotonically to
values that can be considered as slow-diffusion. This is none of those
cases for $\langle T_\mathrm{D}(\mathJr)\rangle$ which decreases
regularly from values similar to $\langle
T_\mathrm{D}(\mathLz)\rangle$ in the centre ($\sim$10~Gyr) to values
around 1~Gyr in the disc. Beyond \UHRb, $\langle
T_\mathrm{D}(\mathJr)\rangle$ shows a plateau between $0.6$ and
$\approx 1$~Gyr which explains the bump around $0.9$~Gyr in
Fig.~\ref{fig:timescales}.  Apart from the very centre, $\langle
T_\mathrm{D}(\mathJr)\rangle$ is always smaller than $\langle
T_\mathrm{D}(\mathLz)\rangle$.

Particles that are retrograde on average (0.1~\mtot, typical of a
barred galaxy, and an identical proportion among never-escaped
particles), have shorter diffusion times for \Lz\ and \Jr\ than for
$E$. For \Lz\ and \Jr\ the values remain compatible with the
timescales in the disk. The scattering of retrograde particles is a
subject in itself because the relative velocity of these particles
with respect to any prograde waves is very large, so that their
interaction can only take place over a very short time. We do not
address this point in this article.

\begin{figure}[htb!]
  \centering
  \includegraphics[keepaspectratio,width=\hsize]{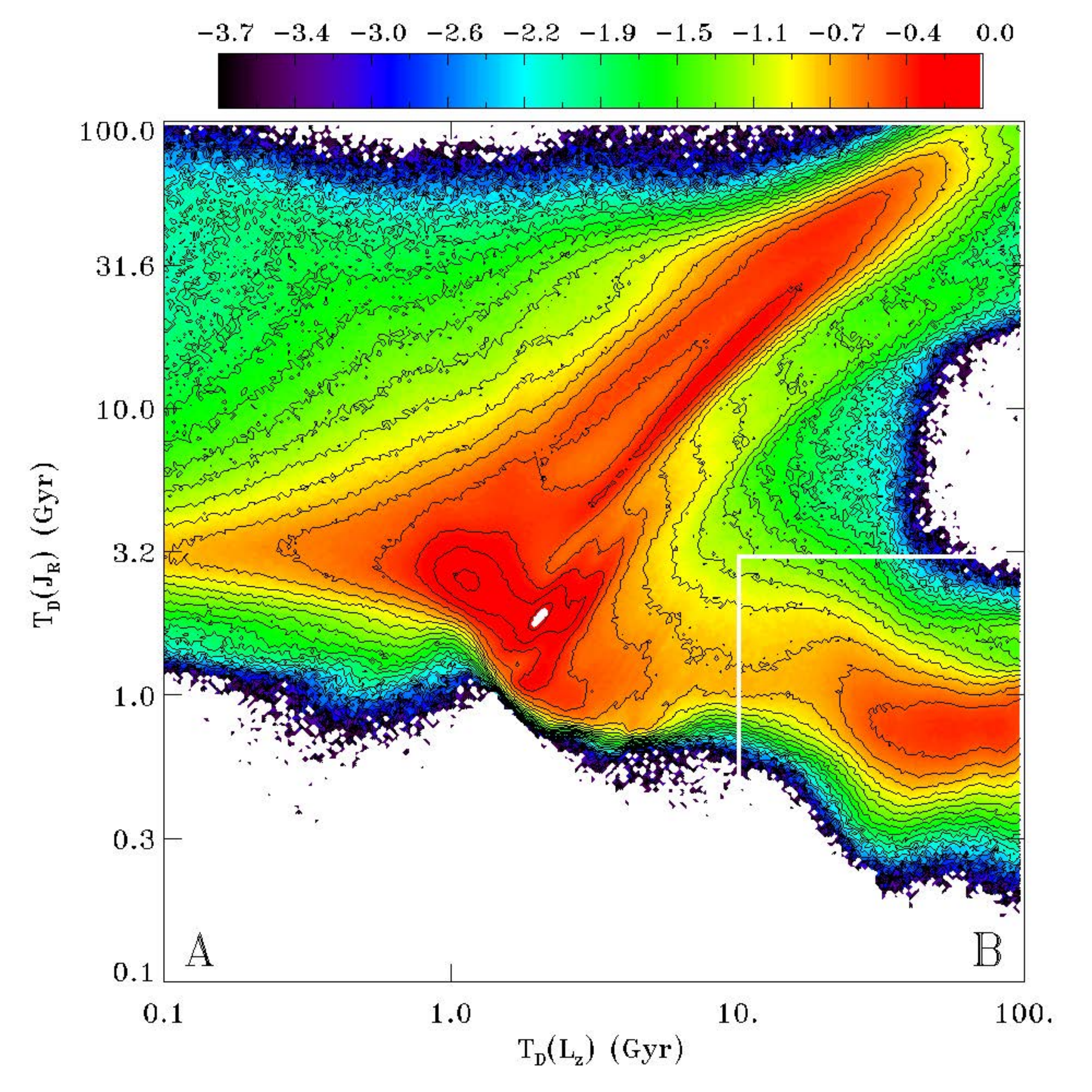}
  \caption{Normalised particle relative frequency (Gyr$^{-2}$) in the
    $T_\mathrm{D}(L_\mathrm{z}) - T_\mathrm{D}(J_\mathrm{R})$ plane in
    $\log$ scale for never-escaped particles only. Contours are spaced
    by 0.15 dex. The white line divides the domain into two sets
    (named \RA\ and \RB), which are used for Fig.~\ref{fig:td_xy} to
    \ref{fig:fd} (see text for details).}
\label{fig:tdlz_tdjr}
\end{figure}

\begin{figure*}[htb!]
  \includegraphics[keepaspectratio,width=\hsize]{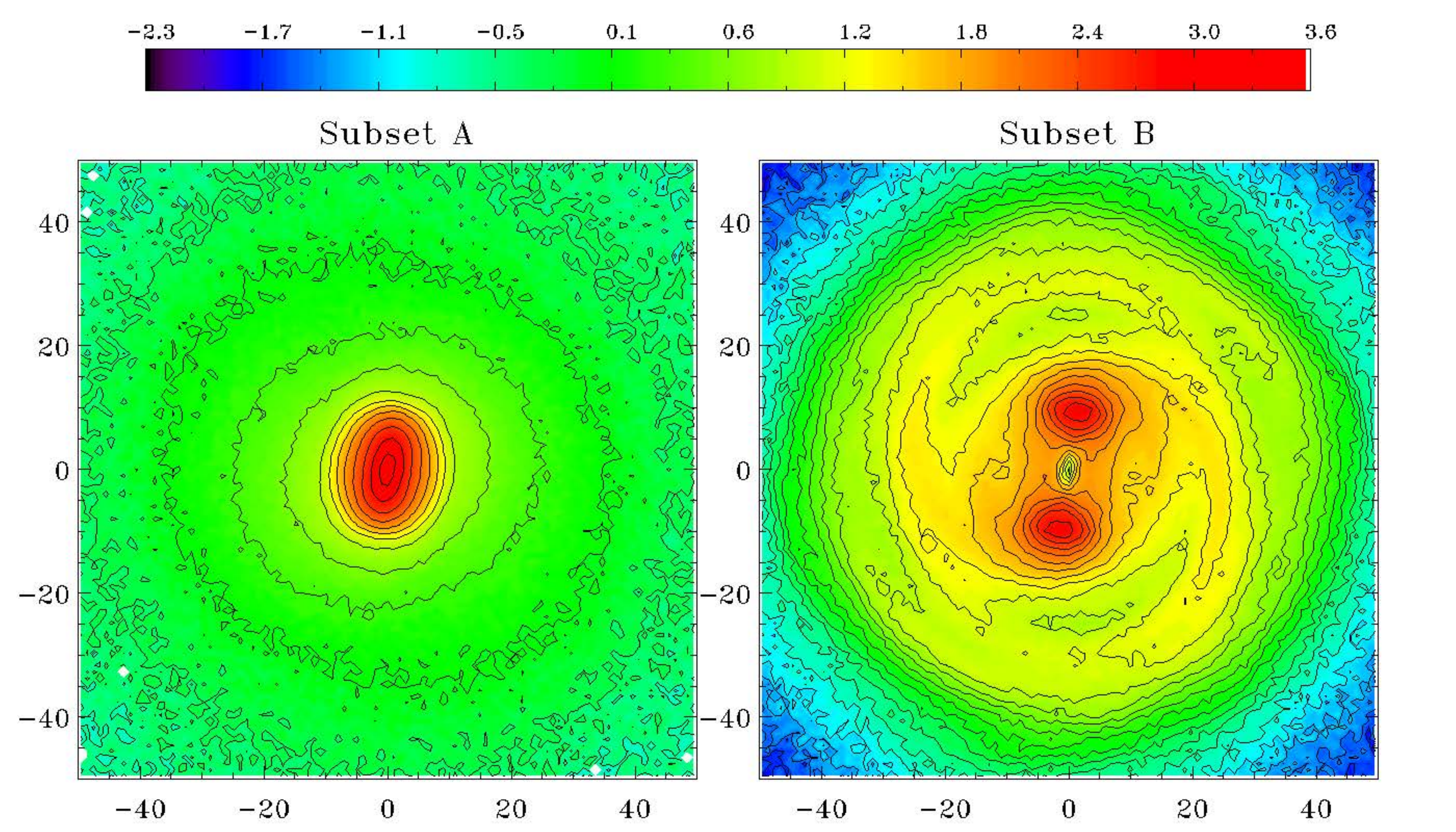}
  \caption{Projected mass surface density (in
    $\mathrm{M}_\sun\,\mathrm{pc}^{-2}$) at $t=10.54$~Gyr inside $\pm
    50$~kpc. Particle sets are defined in Fig.~\ref{fig:tdlz_tdjr} and
    text. The $\log$ colourscale is common to both figures. Black
    isodensities are spaced by 0.4 dex for \RA\ (left panel) and 0.1
    for \RB\ (right panel).
    Spatial scale
    is in kpc.}
\label{fig:td_xy}
\end{figure*}

The particle/mass density distribution in the $T_\mathrm{D}(\mathLz) -
T_\mathrm{D}(\mathJr)$ plane can be studied at different times during
the simulation. Nevertheless, for this first exploratory study, we
found it more interesting to focus on long times. Therefore, this
distribution is analyzed for the final snapshot ($t=10.54$~Gyr) and
plotted in Fig.~\ref{fig:tdlz_tdjr}. It shows many structures. They
correspond to dynamical sets of particles with similar behaviour.
Projected on the $T_\mathrm{D}(\mathJr)$ axis, we can recover the peak
and plateau identified in Fig.~\ref{fig:timescales}.  The most
remarkable structure extends along the bisector
$T_\mathrm{D}(\mathJr)=T_\mathrm{D}(\mathLz)$. It will be seen in the
following that it clearly belongs to the bar. Substructures can be
linked to families of orbits, but this work is beyond the scope of
this article. By integrating along the $T_\mathrm{D}(\mathLz)$ axis,
this large structure explains both the plateau and the quasi-linear
decay observed in Fig.~\ref{fig:timescales} when
$T_\mathrm{D}(\mathJr)> 3$~Gyr. The other noticeable structure, of the
same density and long $T_\mathrm{D}(\mathLz)$ ($>10$~Gyr), is related
to the disc.  It contributes in a large fraction to the peak at
$T_\mathrm{D}(\mathLz))=0.9$~Gyr, although the two structures are
blended by integrating along the $T_\mathrm{D}(\mathLz)$ axis.

It can be expected that these dominant structures correspond to
marked morphological counterparts in physical space. It is quite
difficult to isolate each of the substructures to determine which
morphological element, orbit family or wave type they correspond to.
For the sake of simplicity, we have decided to isolate only one
particular subset (named {\texttt{B}} in Fig.~\ref{fig:tdlz_tdjr} and
\RB\ in the text). After some trials, and focusing on the disc
properties, \RB\ has been defined as $T_\mathrm{D}(\mathLz) > 10$~Gyr
(as in \citet{2020ApJ...889...81W}) and $T_\mathrm{D}(\mathJr) <
3$~Gyr (roughly the end of the plateau in Fig.~\ref{fig:timescales}).
Other thresholds would have been possible (and have been tested) but
these ones roughly isolate particles with ``long''
$T_\mathrm{D}(\mathLz)$ and ``short'' $T_\mathrm{D}(\mathJr)$.
Mass fractions are calculated with respect to the total mass of \PFAA:
0.65~\mtot\ for \RA, 0.24~\mtot\ for \RB, the rest being excluded
particles.

Fig.~\ref{fig:td_xy} shows the mass surface density projected in the
x$-$y plane for the two particle sets defined above. \RA, the most
massive (0.65~\mtot), contains mainly particles with a highly
symmetrical mass distribution, especially the stellar bar. Beyond the
corotation, the distribution in the disc shows only very small
deviations from axisymmetry.  Inside \UHRb\ the morphology is
elliptical-like, which is the signature of strong bars
\citep{2002MNRAS.333..847S,2006A&A...452...97M}.
The case of \RB\ is significantly different. When
$T_\mathrm{D}(\mathLz)>10$~Gyr and $T_\mathrm{D}(\mathJr) < 3$~Gyr,
the mass distribution (0.24~\mtot) shows many morphological
substructures associated to the presence of waves and resonances, such
as spiral arms and rings.  Several structures
exist also inside the bar \citep[such as ansea,][]{2019MNRAS.488..590B}
but extend well beyond \OLRb. As with \RA, there appear to be several
sub-populations, which do not have been separated because of our
approximate criterion for defining \RB.  However, since
\RB\ corresponds to a physical region involved in stellar migration,
we will analyze it in greater detail.

\section{Distribution functions (DF)}
\label{sec:df}

\subsection{DF in \Lz\ and \Jr}
\label{sec:meandf}

\begin{figure}[htb!]
  \centering
  \includegraphics[keepaspectratio,width=\hsize]{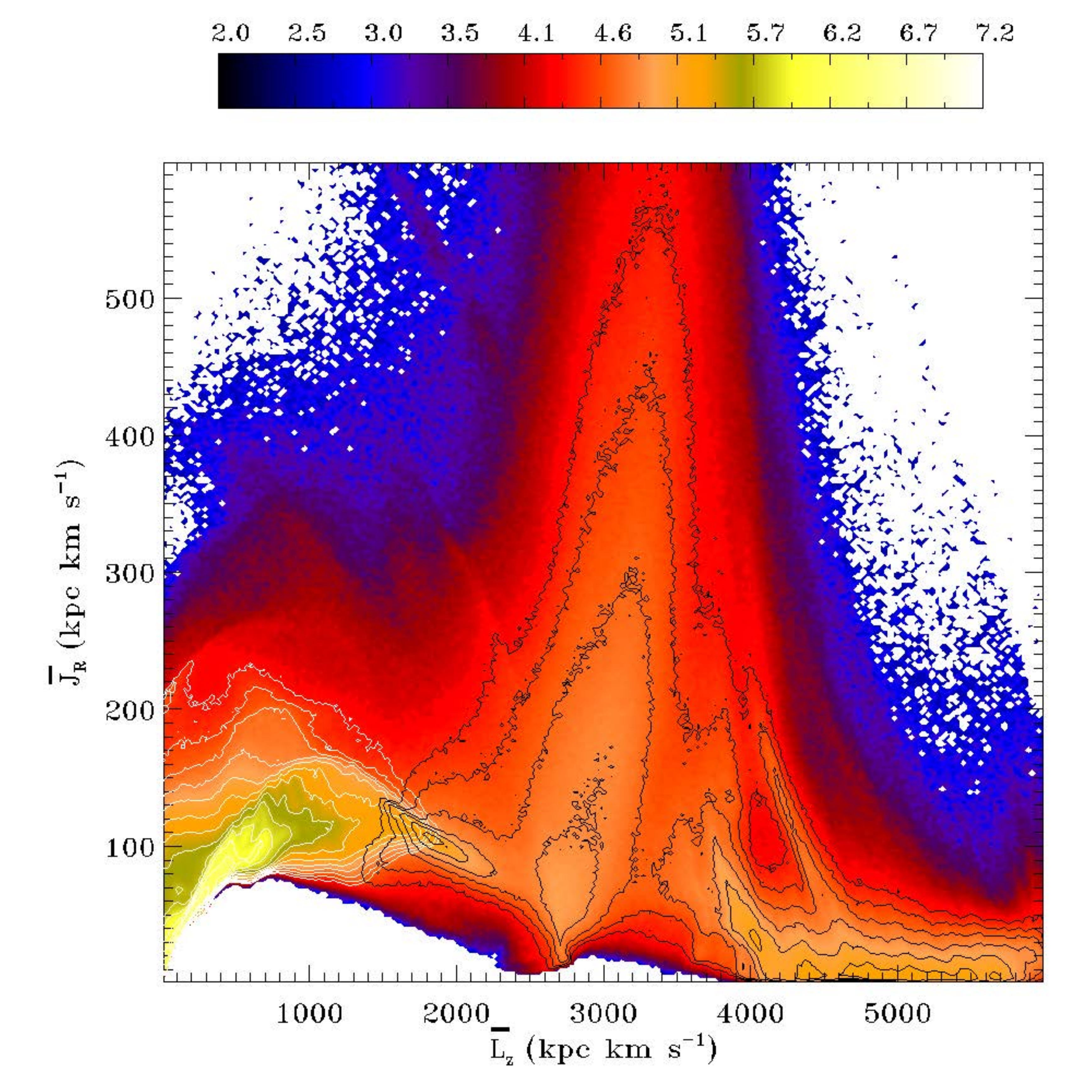}
  \caption{Mass density per (\Lzunit)$^2$ in the
    $\overline{L_\mathrm{z}}-\overline{J_\mathrm{R}}$ plane in $\log$
    scale for never-escaped particles only. This is a time-averaged 2D
    representation of DF$(\mathoLz,\mathoJr)$ over $7.4$~Gyr. Contours
    are spaced by 0.2 dex. White isodensity contours represent \RA,
    black ones \RB.}
\label{fig:lz_jr}
\end{figure}

In Fig.~\ref{fig:lz_jr}, the DF for `never-escaped' particles and the
two particular selections (\RA\, and \RB) are plotted as a function of
$\overline{\mathLz}$ and $\overline{\mathJr}$ time-averaged between
$t=3.16$ and $10.54$~Gyr. The `averaged' DF include the signatures of
all temporal events. This can be compared to
\citet[][]{2012ApJ...751...44S} for instance. However, we should not
expect to find exactly the same results as our initial stellar disc is
not such a stable disc. DF of \PFAA\ is obviously much more structured
than in Sellwood's experiments. In particular, both \Lz\ and \Jr\ bear
the stigma of the bar and its formation (occurring during the first
Gyr). Both integrals have been largely redistributed, especially in
\RA.

\RB\ is identifiable through several substructures. A density peak is
present for $\mathoJr\approx 0$ (circular orbits) and $\mathoLz >
3600$ (beyond \OLRb\ at $t=10.54$~Gyr). This region is bordered by
two almost vertical tails (centered at $4000$ and $4500$~\Lzunit)
which, by similarity to \citet{2012ApJ...751...44S}, may have been
formed by resonant scattering at a Lindblad resonance. That assumption
has to be challenged but we can already claim that it cannot be a
resonance with the bar in this region of the disc.

Between 2300 and 3600~\Lzunit, the large vertical tail with high
\oJr\ values seems to include a significant fraction of the so-called
`hot' population \citep{1987MNRAS.225..653S}. These orbits spend most
of their time outside the bar and sometimes enter inside the bar from
the $L_{1,2}$ Lagrangian points. Tail width is likely to be related to
corotation shifts over time. \oJr\ can reach very high values there.
The mass distribution is very sensitive to how \oJr\ is
calculated. Using the $\overline{\mathEr}/\overline{\kappa}$ ratio as
an estimator of \oJr, the mass would have been concentrated around a
maximum density located at $(\mathoLz\approx 3100, \mathoJr\approx
55)$, which is far from being representative of typical trajectories
in this region. Indeed, particles of the `hot' population explore
large portions of the disc, resulting in large variations in $\kappa$
for each of them, as well as radial kinetic energy \Er. Therefore,
this invalidates the epicycle approximation for `hot' population
orbits.

For $1400 \la \mathoLz \la 2300$~\Lzunit, \RB\ exhibits a ridge that
bridges \RB\ and \RA. This part is linked to ansae identified in
Fig.~\ref{fig:td_xy} and likely associated to \UHRb. A component
separation based on $T_\mathrm{D}$ alone is not identical to a
separation based on morphological criteria. So it is not surprising
that a fraction of \RB\ belongs to what we identify as the stellar
bar. A slightly smarter component separation algorithm could probably
separate this contribution, which seems mainly due to the bar, from
the rest of the disc.

\subsection{DF time evolution}
\label{sec:dfevol}

\begin{figure*}[htb!]
  \centering
  \includegraphics[draft=false,keepaspectratio,width=\hsize]{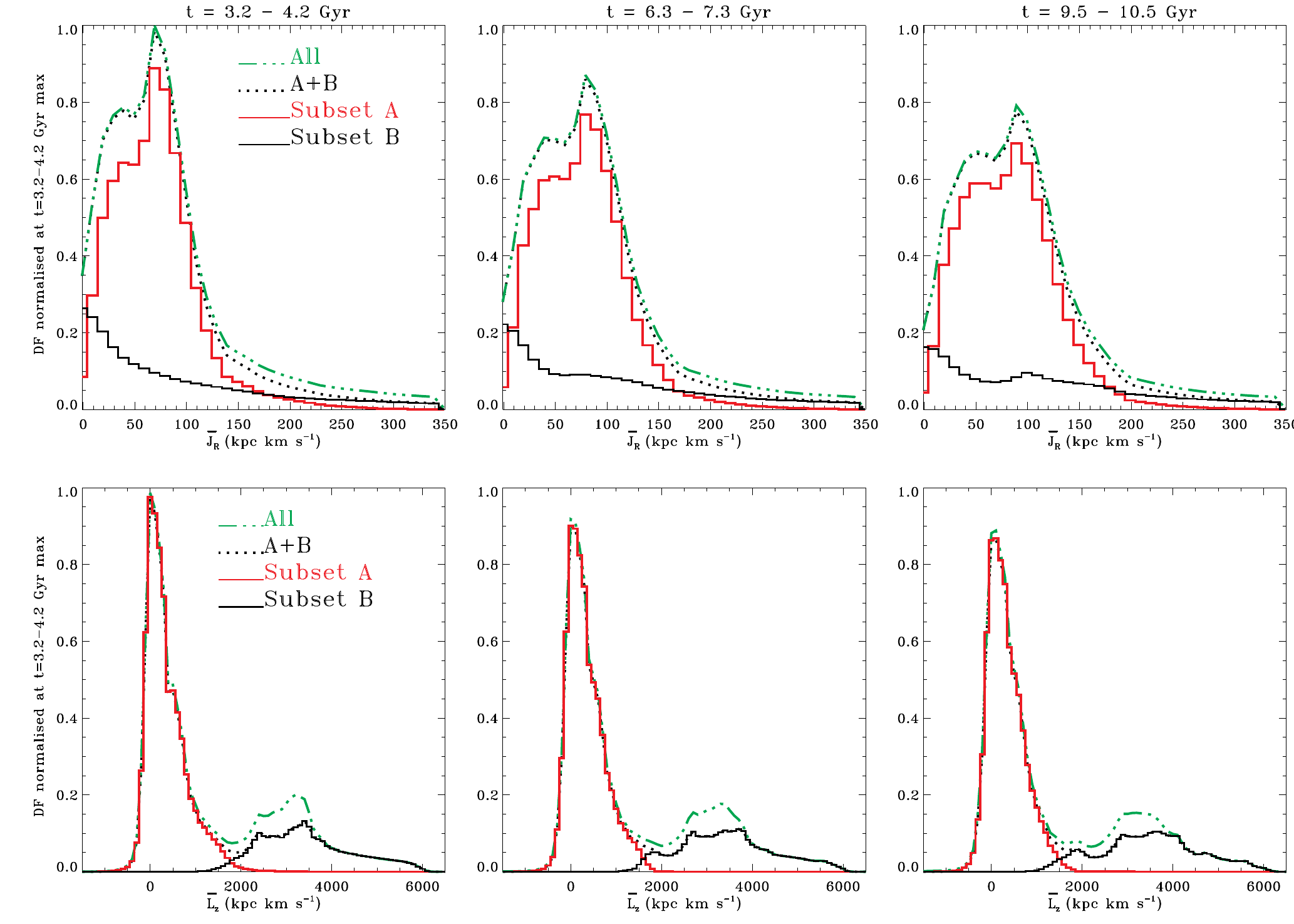}
  \caption{Distribution functions (DF) for all particles (green triple
    dotted-dashed lines), `never-escaped' particles (dotted lines),
    and the two sub-populations defined in Fig.~\ref{fig:td_xy} (red
    and black lines) as a function of \oJr\ (top) and
    $\overline{\mathLz}$ (bottom). \oJr\ and \oLz\ have been averaged
    over $\approx 1$~Gyr, starting at $t=3.2$ (left), 6.3 (middle) and
    9.5~Gyr (right).  DFs have been normalised to
    resp. DF$(\overline{\mathJr}=0)$ and DF$(\overline{\mathLz}=0)$
    for the interval $t=3.2-4.2$~Gyr.}
  \label{fig:fd}
\end{figure*}

1D DF$(\mathJr)$ or DF$(\mathLz)$ can be recovered by the projection of
the $\overline{L_\mathrm{z}}-\overline{J_\mathrm{R}}$ density map on
the axes. DF$(\mathLz)$ is then similar to typical profiles obtained
by a wealth of 3D $N-$body simulations, e.g. those of
\citet{1978ApJ...226..521Z}, \citet{1987MNRAS.225..653S} or
\citet{1991A&A...252...75P}.  The typical DF profile of barred
galaxies has been explained by a superposition of various families of
orbits \citep{1987MNRAS.225..653S,1997A&A...317...14W}, including the
above-mentioned `hot' population.

In order to identify the time evolution of some identifiable
structures in DF$(\mathLz)$, we have calculated \oLz\ over only 1~Gyr
at three different moments of the simulation. The intervals were
centred at $\overline{t}=3.7$, $6.8$ and $10.0$~Gyr. The criteria used
to define the selection of \RA, \RB, and `never-escaped' particles
remain identical as at $t=10.54$~Gyr. For the sake of comparisons, we
have normalized all DF$(\mathLz)$ to the maximum DF$(\mathLz=0)$ at
$\overline{t}=3.7$~Gyr.

In Fig.~\ref{fig:fd} (bottom panels), the contribution of the two
subsets is clearly separated. The whole stellar bar forms the peak of
DF$(\mathoLz)$ and contributes mainly to \RA. The disc, both in a
large axisymmetric fraction and the whole resonant structures, forms
\RB. This region contains the `hot' population bump
($\overline{L_\mathrm{z}}\ga$~2300~\Lzunit).  Unsurprisingly,
particles that spend some time outside the simulation grid
preferentially belong to the disc. They make a significant
contribution to the `hot' population bump. The secondary smaller bump
for $\overline{L_\mathrm{z}}\la$~2500~\Lzunit\ has been
previously identified as a ridge overlapping with \RA. It is linked to
bar structures (cf. Fig.~\ref{fig:td_xy}).

Other evidence, the number of particles with $\mathLz=0$ decreases
over time which means that the number of particles close to the
centre or on radial orbits decreases. As the particle number of
\RA\ is constant by definition, this also means that the
redistribution of \Lz\ within the bar continues. For \RB, resonant
structures are visible as small oscillations along the `hot'
population bump.

The projected \oJr\ distribution (Fig.~\ref{fig:fd} top panels) blends
all the components discussed so far.  Most particles temporarily
escaped have $\mathoJr \ga 150$~\Lzunit. The mode is mainly due to
\RA. Its position shifts from $\overline{J_\mathrm{R}} \approx
70$~\Lzunit\ at $\overline{t}=3.7$~Gyr to $\approx 90$~\Lzunit at
$\overline{t}=10.0$~Gyr. In this timeframe, the distribution width
increases by $\approx 23$\%\ for \RA\ while a substructure appears in
the \RB\ distribution, leading to a small bump around $\mathoJr\approx
100$~\Lzunit. This behaviour is symptomatic of moderate but regular
radial heating of the disc.

\section{Evolution of circular orbits ($\mathJr=0$)}
\label{sec:zeroevol}

\begin{figure}[htb!]
  \centering
  \includegraphics[draft=false,keepaspectratio,width=\hsize]{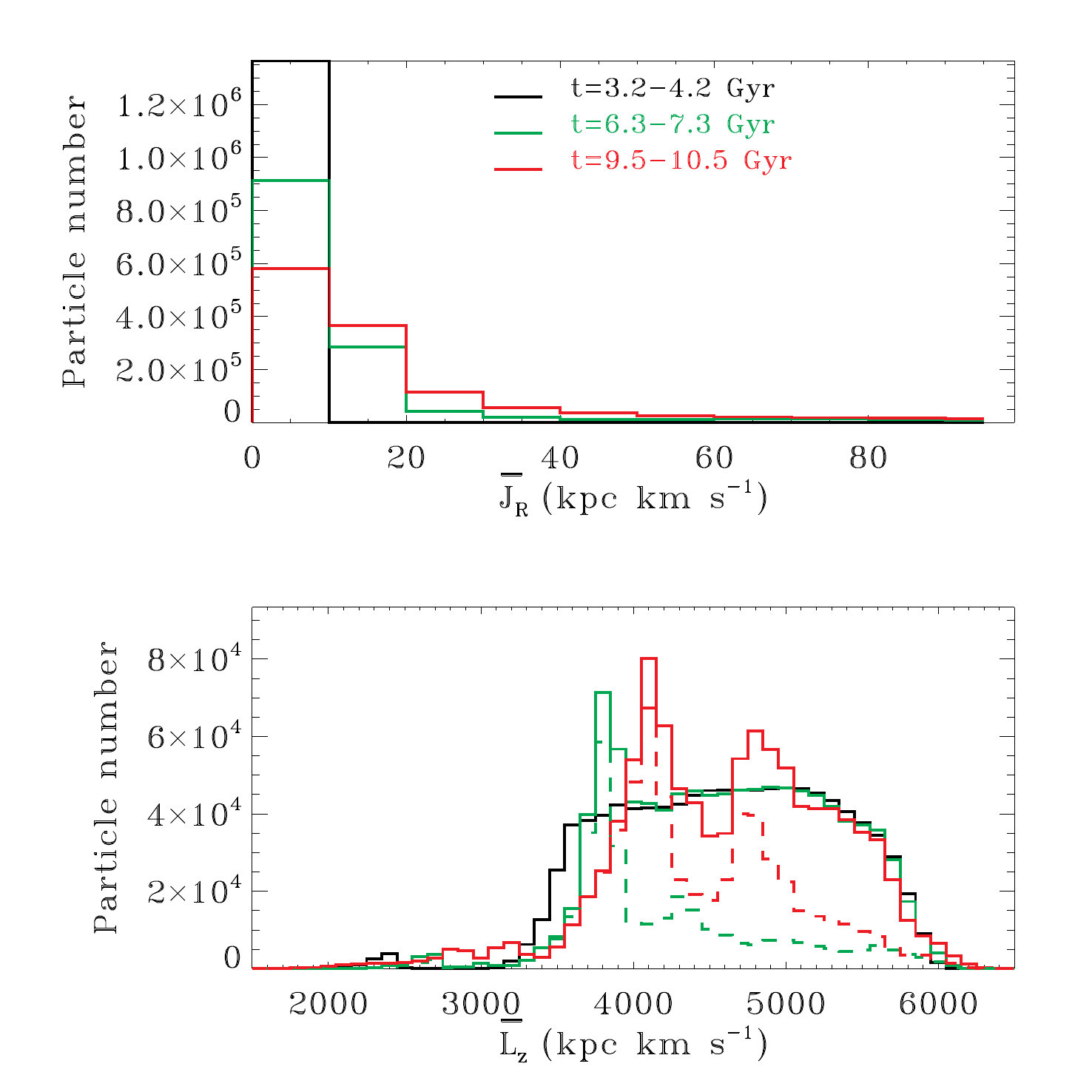}
  \caption{Particle number as a function of \oJr\ (top) and
    \oLz\ (bottom), for particles selected as $\mathoJr(3.2-4.2) <
    10$~\Lzunit\ (black lines). The same particle selection is drawn
    at $t=6.3-7.3$ (in green) and $t=9.5-10.5$~Gyr (in red). The
    contribution of particles with strictly increasing \Jr\ to
    DF(\oLz) is mentioned by a dashed line.}
  \label{fig:jr0}
\end{figure}

\begin{figure*}[htb!]
  \centering
  \includegraphics[draft=false,keepaspectratio,width=\hsize]{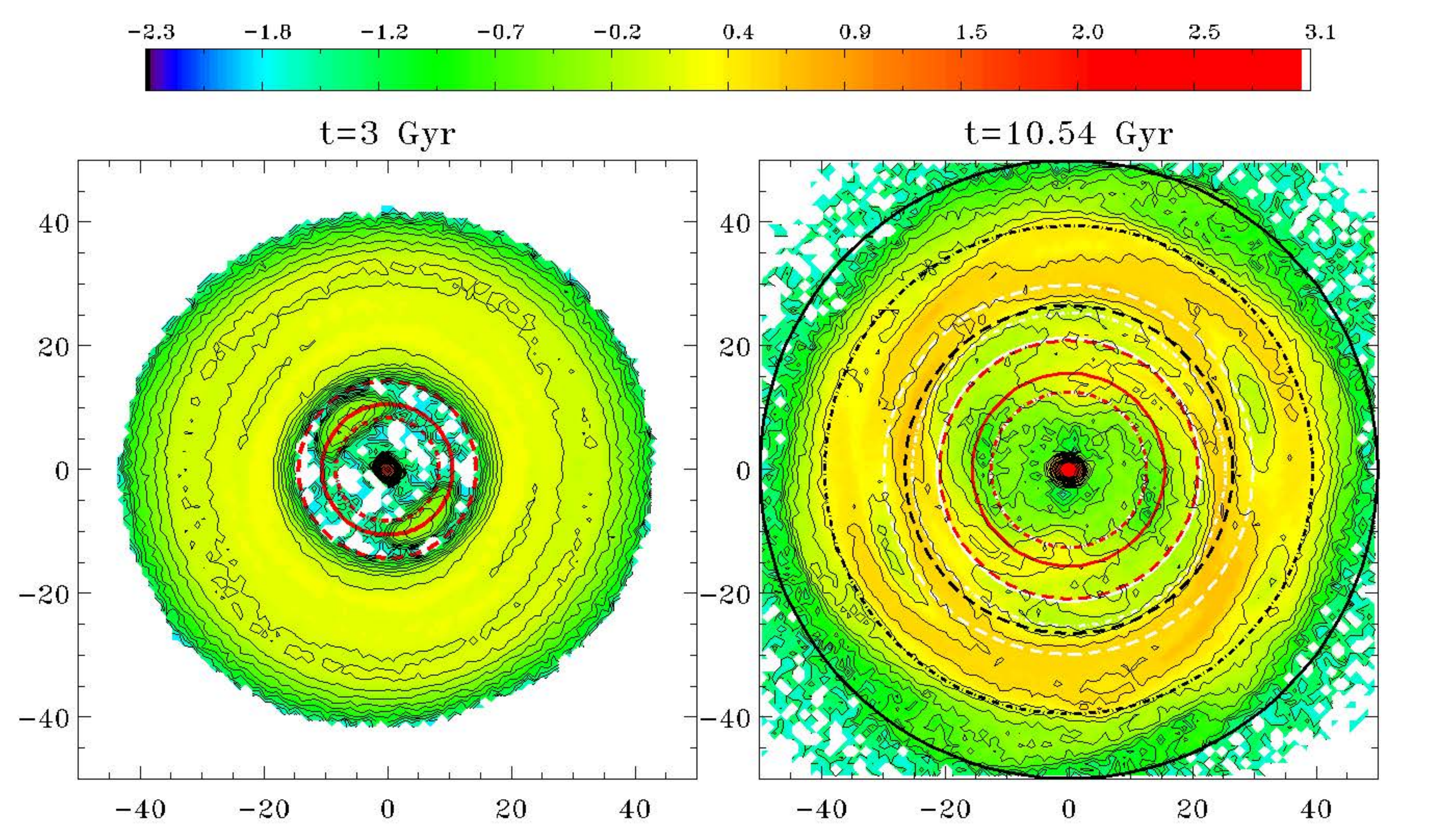}
  \caption{Projected mass surface density at for particles selected as
    $\mathoJr(3.2-4.2) < 10$~\Lzunit\ at $t=3$~Gyr (left panel) inside
    $\pm 50$~kpc. Their distribution at $t=10.54$~Gyr (right panel).
    The $\log$ colourscale is common to both figures (in
    $\mathrm{M}_\sun\,\mathrm{pc}^{-2}$). Black isodensities are
    spaced by 0.2 dex.  Circles show the position for ILR (dot-dashed
    line), UHR and outer $m\!=\!+4$ resonance (short-dashed), CR (full
    line), and OLR (long-dashed).  In red, resonances for the stellar
    bar: \UHRb, \CRb, and \OLRb, as defined in
    Sect.~\ref{sec:wave}. Using the same linestyle, the white circles
    represent the intermediate spiral: \ILRi\ (close to \UHRb),
    \CRi\ (close to \OLRb), outer $m\!=\!+4$ resonance and \OLRi. The
    black circles represent the outer wave \ILRs (close to
    intermediate spiral outer $m\!=\!+4$ resonance), \UHRs, and \CRs.  }
  \label{fig:xy0}
\end{figure*}

Let us take a closer look at what may be one of the causes of radial
heating in \PFAA. The spread of DF$(\mathoJr)$ increases over time as
the bin $\mathoJr\approx 0$ depopulates. Since DF$(\mathoJr)$ has been
normalised by DF$(\mathoJr=0)$ at $\overline{t}=3.7$~Gyr,
Fig.~\ref{fig:fd} clearly shows that the number of particles with
$\mathoJr\approx 0$ decreases with time, even long after the bar has
been formed.  This strongly points to an increase in epicycle
amplitude as $\mathJr \sim \frac{1}{2}\kappa X^2$. The evolution of
these particles on near-circular orbits thus deserves a particular
analysis since the variation of \Jr\ is discriminating with respect to
the blurring/churning issue.

In order to take the numerical uncertainties inherent to this type of
simulation into account, let us define hereafter circular orbits as
$\overline{J_\mathrm{R}} < 10$~\Lzunit.  If we select only the
particles with $\mathoJr(\overline{t}=3.7) < 10$, the evolution of
DF$(\mathoJr)$ and DF$(\mathoLz)$ can be extracted at time
$\overline{t}=6.8$ and $\overline{t}=10.0$~Gyr
(Fig.~\ref{fig:jr0}). This circular orbit population contains about
0.039~\mtot\ (i.e. $7.7 \times 10^9$~\msol). Almost all particles are
located in the \oLz\ tail in the outermost part of the disc ($\mathoLz
\gg 3300$), and mostly well beyond \OLRb\ (located at 20.8~kpc,
i.e. $\mathLz\approx 3790$~\Lzunit, at $t=10.54$~Gyr).  Very few of
these particles belong to the `hot' population bump which is centred
on $\mathoLz\approx 3000$.

Fig.~\ref{fig:jr0} shows that \oJr\ increases significantly for
$\approx 57.4$\%\ of particles initially on near-circular orbits. It
leads to an increase of $\sigma_\mathrm{R}$ of the order of $\approx
10$~\kms\ beyond \OLRb. This is comparable with observations of solar
neighbourhood \citep[][for instance]{2008A&A...480...91S,
  2019MNRAS.489..176M}, although \PFAA\ does not pretend to reproduce
the properties of Milky-Way.  The corresponding impact on
DF$(\mathoLz)$ is perfectly identifiable by spikes in the bump of this
population. This is clearly the signature of a coherent mode of
\Lz\ exchange, coupled with the increase in \Jr, presumably in the
form of waves. Well beyond \OLRb, it is clearly the impact of
propagating density waves with $\Omega_\mathrm{p} < \Omega_\mathrm{B}$
that is illustrated here. According to Eq.~\ref{eq:djdlde}, if the
particle scattering is resonant with a wave, it cannot be at
corotation since $\Delta\mathJr\neq 0$.

The mass surface density at $t=3$~Gyr of particles selected at
$\overline{t}=3.7$~Gyr as having $\mathoJr\approx 0$ is plotted in
Fig.~\ref{fig:xy0}. As expected, this set of particles has a very
axisymmetric distribution, mostly beyond \OLRb, with the notable
exception of those inside the bar, which are concentrated around the
Lagrange L$_{4,5}$ points (the bar is oriented at about 45\degr).  At
$t=10.54$~Gyr, the distribution of the same particle set shows many
wavelet-type structures. \Jr\ of more than half of the particles
having increased significantly, it is normal that the initial
axisymmetry has been broken. In anticipation of Sect.~\ref{sec:wave},
we have plotted some resonances identified from three patterns
detected in the disk: the bar, an intermediate spiral structure, and a
set of external waves. The structures that appear progressively
between $t=3$ and $t=10.54$~Gyr are mainly beyond \OLRb\ and do not
exceed the corotation of the outermost waves.

\section{Wave analysis}
\label{sec:wave}

\subsection{Fourier spectrograms}
\label{sec:spectrograms}

\begin{figure*}[htb!]
  \centering
  \includegraphics[draft=false,keepaspectratio,width=0.33\hsize]{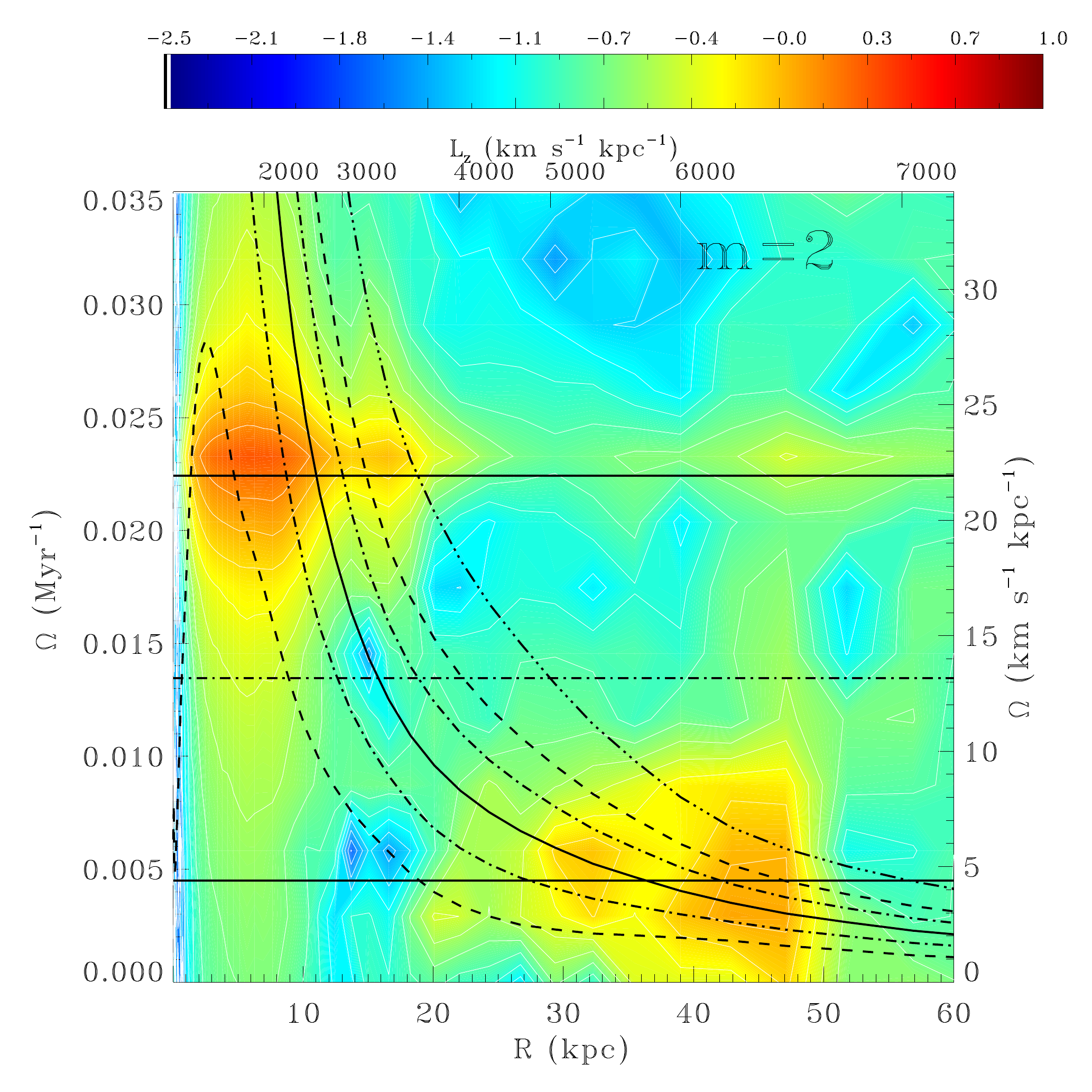}
  \includegraphics[draft=false,keepaspectratio,width=0.33\hsize]{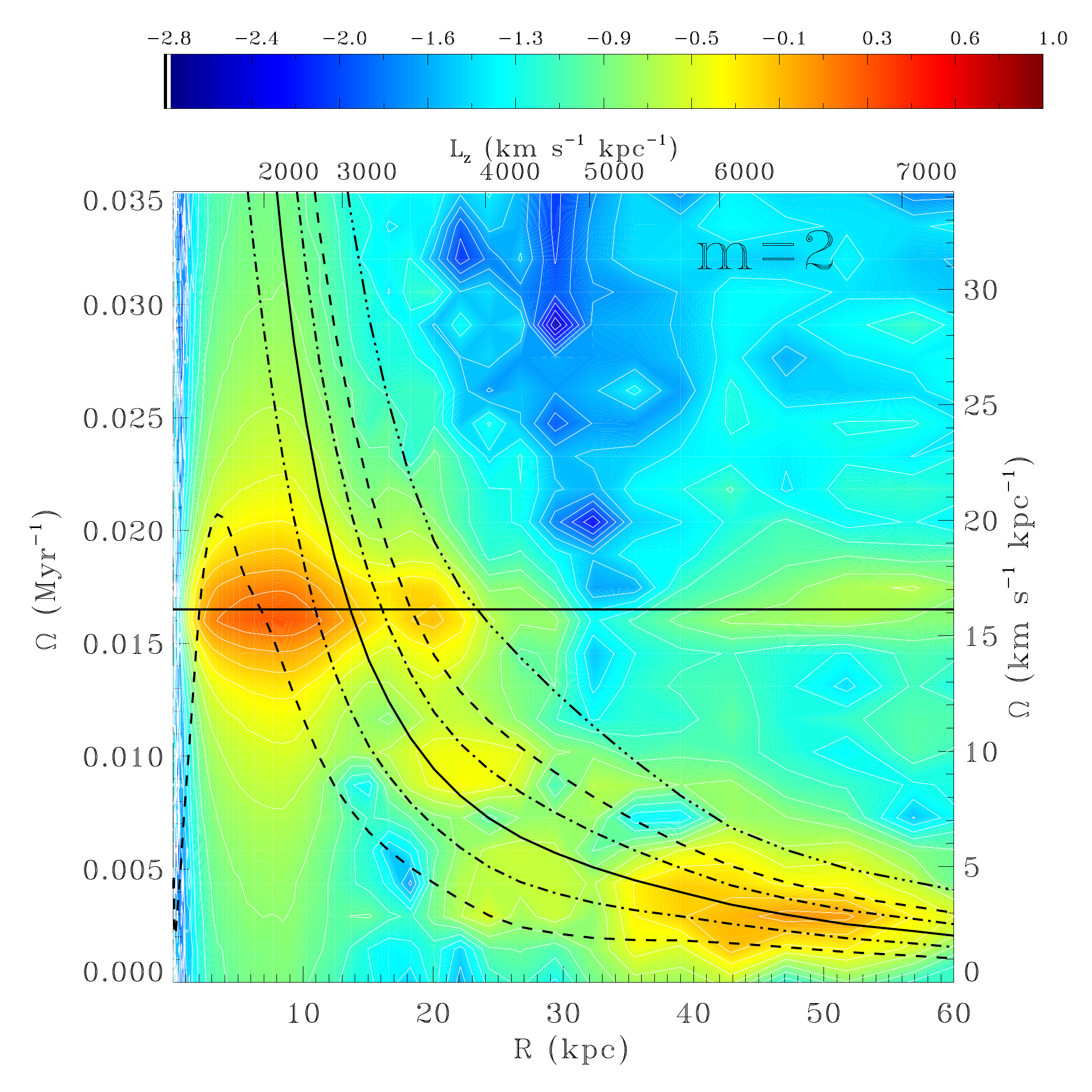}
  \includegraphics[draft=false,keepaspectratio,width=0.33\hsize]{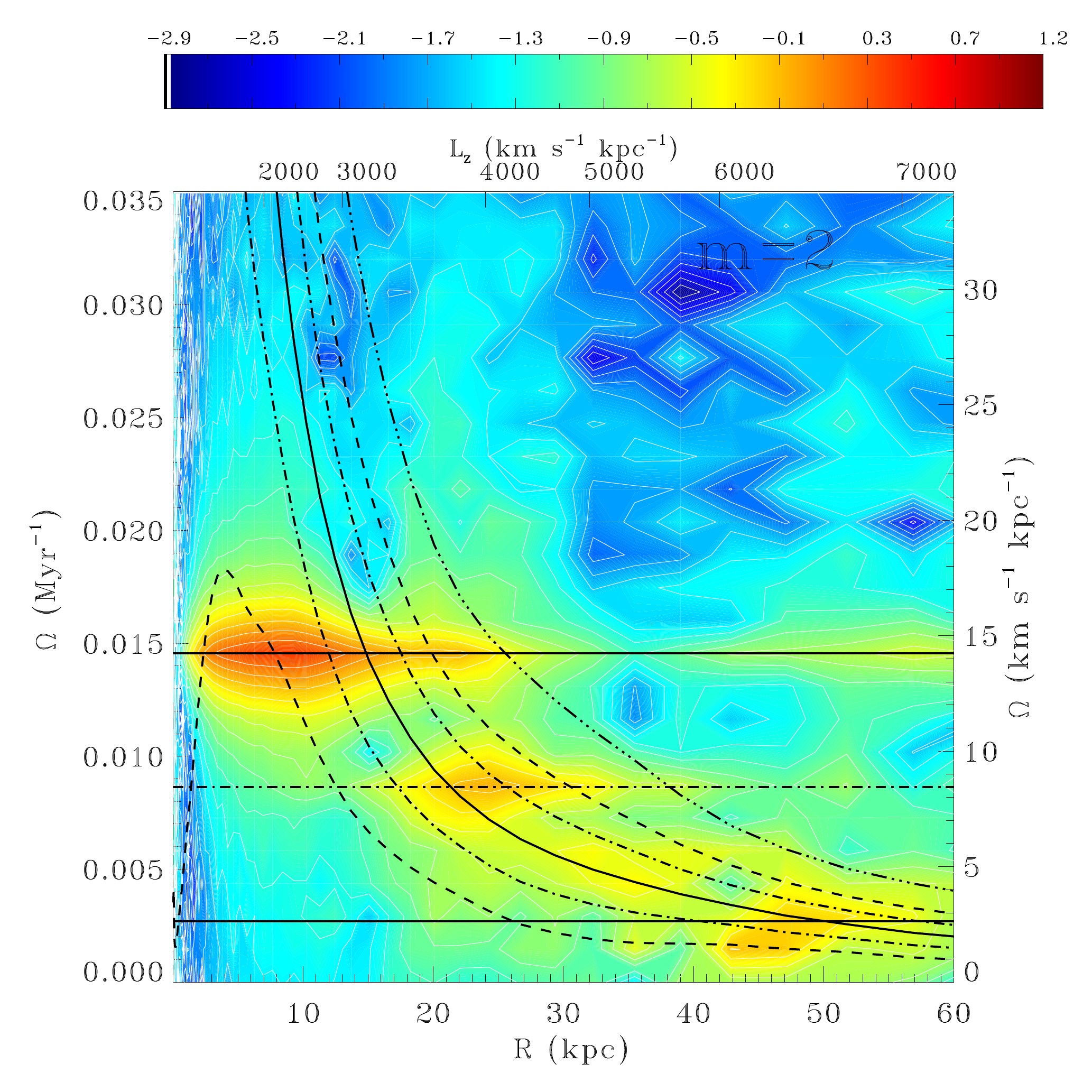}
  \includegraphics[draft=false,keepaspectratio,width=0.33\hsize]{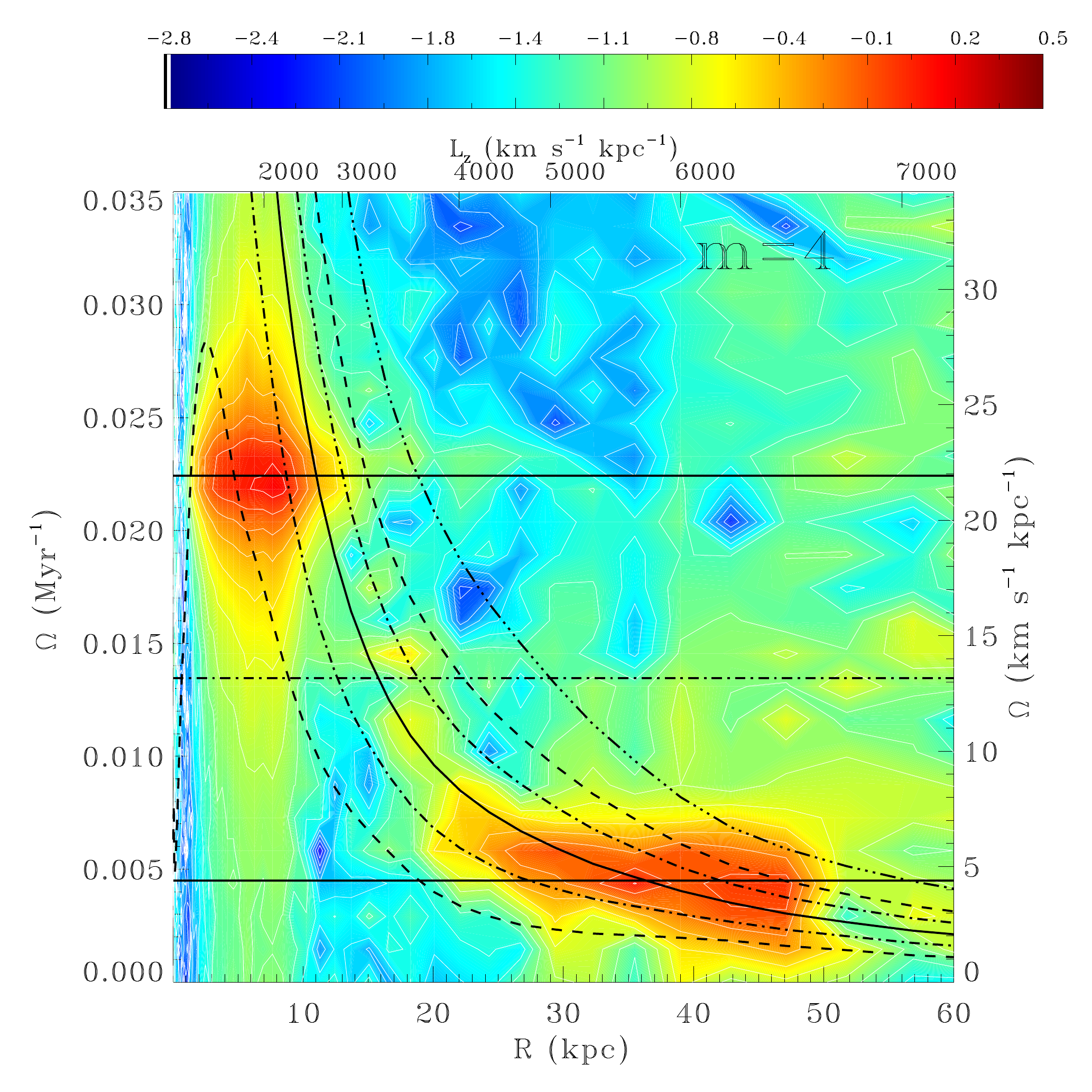}
  \includegraphics[draft=false,keepaspectratio,width=0.33\hsize]{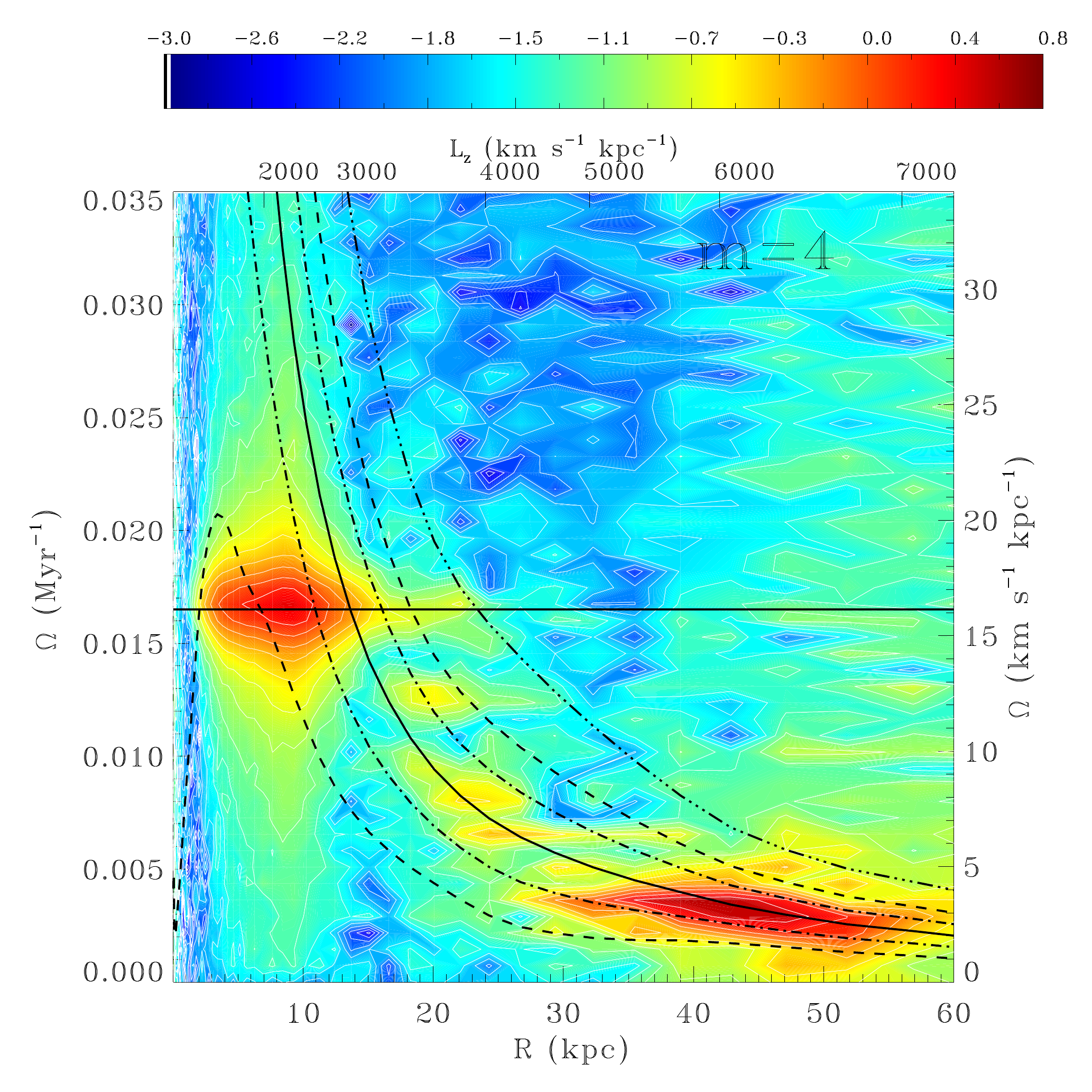}
  \includegraphics[draft=false,keepaspectratio,width=0.33\hsize]{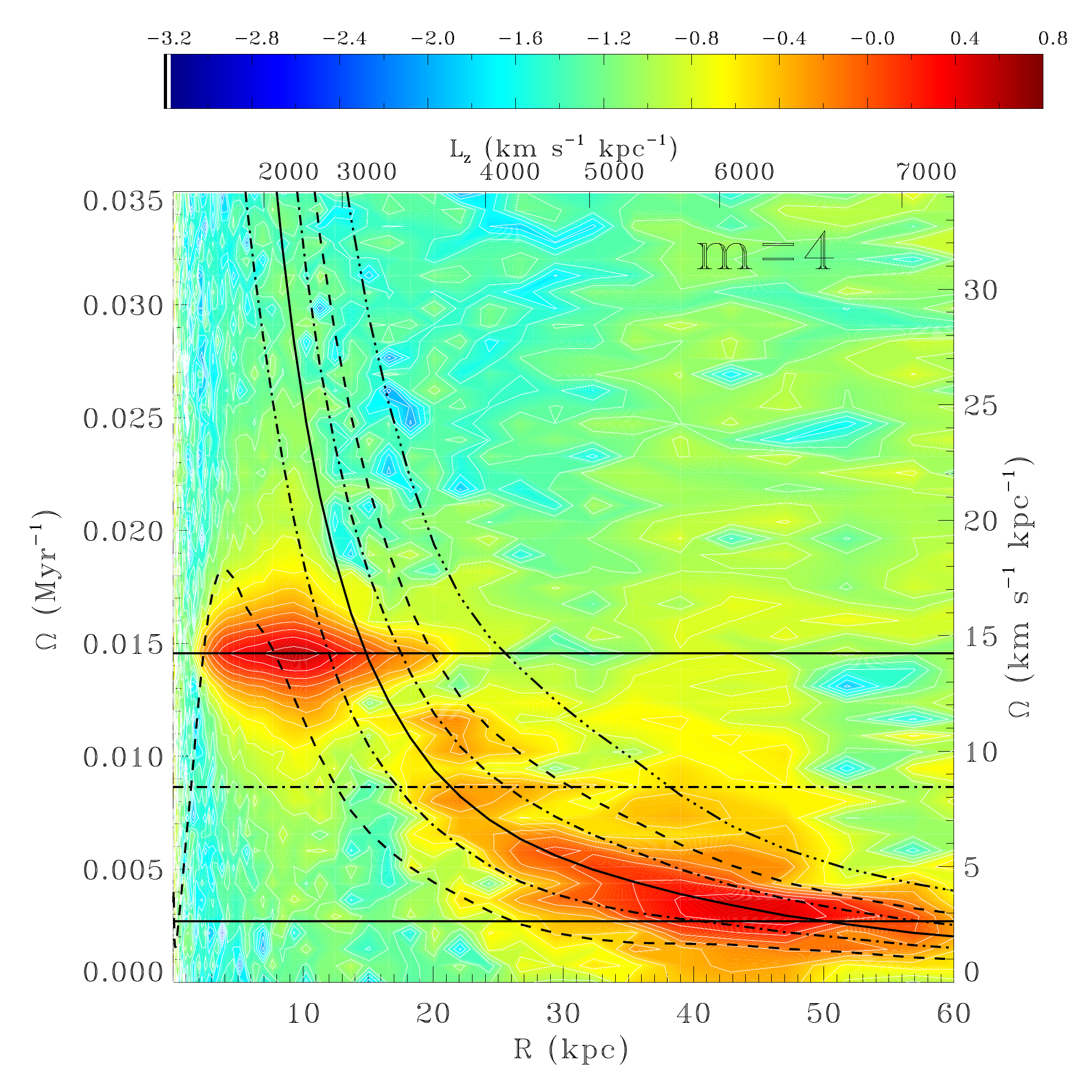}
  \caption{Top row: $m=2$ power spectra in $\log$ scale as a function
    of radius for \PFAA. The time windows are $3.16-4.24$~Gyr (left
    panel), $6.32-8.48$~Gyr (middle) and $8.38-10.54$~Gyr (right). The
    vertical scales give values of $\Omega$ in Myr$^{-1}$ (left) and
    in \omegaunit\ (right). The radial scale (in kpc) is converted in
    \Lz\ using the circular rotation curve. The mean curves $\Omega
    \pm\kappa/2$ are drawn as black short-dashed lines, $\Omega
    \pm\kappa/4$ as dot-dashed lines, $\Omega$ as a solid line (for
    the CR), and $\Omega+\kappa$ as a triple-dot-dashed line. The
    uppermost full horizontal line represents the mean bar pattern
    speed $\Omega_\mathrm{B}=21.9-14.2$ (full line) determined
    directly from the time variation of bar position-angle. The lowest
    one is an estimated intermittent waves packet
    $\Omega_\mathrm{oW}\approx 4.4-2.6$~\omegaunit. The intermediate
    wave at $\Omega_\mathrm{iS}\approx 13.2-8.4$~\omegaunit\ is
    computed as a beat mode.  Bottom row: same for $m=4$.  }
  \label{fig:fft}
\end{figure*}

In order to give a coherent overview of the dynamical mechanisms at
work in this simulation, we have been looking for waves in the disc
that could be associated with variations in angular momentum and
radial action.  Fig.~\ref{fig:fft} show the classical $m=2$ and $m=4$
spectrograms cumulated in time windows $t=3.16-4.24$, $6.32-8.48$ and
$8.38-10.54$~Gyr, respectively. The first window (1075~Myr wide) is
shorter than the others (twice as large) as the slowdown rate of
$\Omega_\mathrm{B}$ is higher when the bar is young.

We have identified at least three kind of waves that might imprint the
evolution of the disc in a significant way:

\begin{itemize}
\item{The mode identified as the bar covers a large domain in
  $\Omega_\mathrm{p}=\Omega_\mathrm{B}$, from $23.2$ down to
  $13.5$~\omegaunit. Since the integration time window is large, and
  the bar is secularly slowing down, it is normal that this mode is
  spread between the extreme values of $\Omega_\mathrm{B}$.  Beyond
  the bar corotation (\CRb), the bar permanently excites a $m=2$ mode
  (thus of identical speed pattern), visible beyond the bar
  corotation, but also well beyond the 1:1 resonance
  ($\Omega+\kappa=\Omega_\mathrm{B}$ resonance).}

\item{An intermediate spiral wave (named `iS' hereafter), whose
  maximum power is located at $\Omega_\mathrm{iS} \in
  [15-8]$~\omegaunit. This mode has a spatial extension that goes
  roughly from \UHRb\ to its own $1/1$ resonance, i.e. between
  $\approx 20$ and $\approx 40$~kpc (cf. also
  Fig.~\ref{fig:xy0}). This mode gains power over time. It is clearly
  more visible at the end of the simulation, but seems to be present
  as soon as $t=3$~Gyr.}

\item{At lower values of $\Omega$ (below $\approx 10$~\omegaunit),
  other waves appear, which are not permanent. They reappear regularly
  at slightly decreasing $\Omega_\mathrm{p}$ values, giving an average
  contribution that is spread out in $\Omega$. Nevertheless, cumulated
  over $7.4$~Gyr, their signature (in terms of power) is at least
  equivalent to the bar. The behaviour is similar to wave packets that
  carry angular momentum outwards in a finite time and not like
  standing waves. The integration over about 2 Gyr shows a cumulative
  power that exaggerates the comparison with the stellar bar, which is
  permanent. These wave packets are nevertheless indispensable to
  evacuate the angular momentum towards the outer edge of the disc.
  For the most powerful of these intermittent waves, we have roughly
  determined $\Omega_\mathrm{oW}$ and plotted it in Fig.~\ref{fig:fft}
  ($\Omega_\mathrm{oW}\in [4.4-2.6]$\omegaunit, where 'oW' stands for
  `outer wave'). These values, estimated by hand, are very approximate
  because they are slightly different according to $m$.  We can
  reasonably approximate that this recurrent wave structure, whose
  successive values of $\Omega_\mathrm{oW}$ decrease, is equivalent to
  a permanent wave that would slow down over time, as the bar does.
  This structure is complex and difficult to analyze because it is the
  only one whose power spectrum does not cancel neither for $m>4$ nor
  for odd $m$. Its trace is perfectly visible up to $m=8$, the limit
  we imposed on ourselves in our study, while for $m>4$ the other inner
  structures have almost no contribution.}
\end{itemize}

The nature of the intermediate spiral wave raises question as its
normal mode is very close to that obtained by beating the bar mode
with the averaged outer waves discussed above.  In a linear approach,
all the waves present in the disk evolve independently and do not
interact. However, if higher order terms of kinetic equations are
considered, this is no more true and waves can exchange energy and
angular momentum \citep{1988MNRAS.232..733S}. Selection rules on wave
numbers and frequencies then apply.  Therefore, applying these
selection rules,
$\omega_\mathrm{beat}=2\Omega_\mathrm{B}+2\Omega_\mathrm{oW}$
decreases from
$\approx 53$ to $\approx 34$, giving
$\Omega_\mathrm{iS}=\omega_\mathrm{beat}/4\approx 13.3-8.5$~\omegaunit. This
is approximately the location of the intermediate wave in $m=2$ and $m=4$
spectrograms.  Therefore, this could correspond to the mode coupling
as illustrated by \citet{1997A&A...322..442M} in an N$-$body
simulation. 

\subsection{Resonance overlaps}
\label{sec:overlaps}

With at least three patterns in the stellar disc, a number of
resonance overlap is unavoidable. Over 1~Gyr, resonance radii can
increase by up to $\approx 1$~kpc due to the changes in
$\Omega_\mathrm{p}$ and in $\Omega(R)$ and $\kappa(R)$
curves. Moreover, resonances have unavoidably a width which must be
express in frequency units. For analytical dynamical systems, this
width is often computed in the pendulum approximation and is typically
proportional to the square-root of the perturbation amplitude. For a
galaxy, we can make the reasonable assumption that the width depends
both on $\Omega_\mathrm{p}$ bandwidth and on the local slope of
$\Omega(R)+\kappa(R)/m$.  The latter dependency results in narrower
resonances when only a single bar is involved, the ILR being probably
the narrowest of all, the OLR being the widest. On the other hand,
resonances with spiral structures are much wider. As for the width of
$\Omega_\mathrm{p}$, it depends on its time derivative and therefore
on possible fluctuations.

Therefore, the notion of overlap, expressed in the spatial domain,
should not be taken literally. The above margin of the order of 1~kpc
can be applied. In which case, between $t=6.32$ and $8.48$~Gyr
(Fig.~\ref{fig:fft} middle panel), \OLRb\ and \CRi\ are close to each
other, as well as \UHRb$-$\ILRi.  Between $t=8.38$ and $10.54$~Gyr
(Fig.~\ref{fig:fft} right panel), \UHRb\ is still close to \ILRi,
while \OLRb\ and \CRi\ are now separated by $\approx 1.5$~kpc. Within
the large $\Omega_\mathrm{oW}$ uncertainties, \ILRs\ might also be
close to both the bar $1/1$ resonance and the outer $m\!=\!+4$
intermediate wave resonance.  Between these two extreme time windows,
any other type of overlap may occur.

The bar and the intermediate wave seem to be locked as
\UHRb$-$\ILRi\ overlap is constant within less than 0.5~kpc while
\OLRb$-$\CRi\ do the same within a slightly wider range. If the
intermediate spiral is a beat mode, it would imply that the outer
waves pattern is also locked to the bar one. In view of the
uncertainties in determining the pattern speed of the outer wave, we
cannot firmly confirm this.

The frequency analysis thus suggests that the dynamical particle/wave
interactions have many sources in the disc. This is, in particular,
one of the reasons why we claimed to be unable to confirm that
\OLRb\ might be a barrier to radial migration
\citep{2020ApJ...889...81W}.  Indeed, \OLRb\ occurs at a rather small
radius with respect to the whole disc extension. Any single wave (as a
bar) cannot efficiently carry angular momentum over a large radial
span. Therefore, spiral waves take over the bar in angular momentum
exchanges, at least up to the \UHRs\ of the lowest frequency outer
waves (Fig.~\ref{fig:xy0}).  The \UHRs\ resonance seems to mark the
end of the set of external waves \citep[as for][]{1994A&A...286...46P}
whose properties, both morphological and temporal, are different from
the intermediate spiral.

\section{Discussion}
\label{sec:discussion}

First of all, let us recall that we focus on the evolution of the disc
once the bar formation phase is over, so $3.16 < t < 10.54$~Gyr for
\PFAA. The disc is then in a state of adiabatic evolution. It remains
subject to the gravitational influence of the bar and to its own
self-gravitating instabilities.

The diffusion timescale, in Chirikov's sense, is shorter for \Jr\ than for
\Lz\ in the disc. In average, $T_\mathrm{D} (\mathJr)$ is even
slightly shorter in the disc than in the bar. The set of
\RB\ particles, selected such that $T_\mathrm{D}(\mathJr) < 3$ and
$T_\mathrm{D}(\mathLz) > 10.54$, is associated with all morphological
structures related to the secular evolution of the disc, bar excluded.
In the context of the epicycle approximation, \Jr\ diffusion could be
interpreted as $X$ diffusion and thus secular radial heating. In other
words, blurring \citep{2009MNRAS.396..203S} would be favoured over
churning for \RB\ in the time window $3.16 < t <
10.54$~Gyr. Alternatively, one could also imagine that \Jr\ diffusion
could be more strongly linked to that of $\kappa$. That the $\kappa$
frequency can diffuse is mainly due to its dependence on $R$, which
would translate into \Lz\ diffusion. However, as
\citet{2015A&A...578A..58H} showed, the disc evolution is complex
since blurring and churning coexist and their relative importance
evolves over time. The magnitude of churning decreases with time,
unlike blurring. In our case, starting our analysis well after the bar
formation, the intensity of churning may have strongly decreased.

The stellar disc is not only made of \RB\ particles.  On the one hand
there are disc particles in the complementary subpopulation \RA\ for
which $T_\mathrm{D}(\mathJr)<3$ \emph{and}
$T_\mathrm{D}(\mathLz)<10.54$ (lower-left corner of
Fig.~\ref{fig:tdlz_tdjr}). Thus, a fraction of the disc shows
\Lz\ variations on short timescales, i.e. $< 10.54$~Gyr. It should be
remembered that the numerical values of the boundaries are not yet
firmly established.  On the other hand, the DF of these \RA\ particles
also evolves: both DF(\Jr) and DF(\Lz) widen by $\approx
20$\%\ (according to their FWHM) while their maxima decrease. The
widening of DF(\Lz) for \RA\ is mainly due to an increase in \Lz\ for
particles initially with $\mathLz\approx 0$. This coevolution of
\Jr\ and \Lz\ suggest that the scattering mechanism is also at work in
this simulation, but on a longer timescale compatible with a
decreasing importance. This dynamical mechanism cannot be only the
scattering by corotation(s) as $\Delta\mathJr\neq 0$. Inside the
stellar bar, ILR scattering is likely the most efficient mechanism.

In the framework of a particle-mesh code, the wave$-$particle
exchanges shape the evolution of dynamical properties. Several waves
have been identified by their power spectrum. Some are highly
time-dependent. Intermittance has not been studied exhaustively yet
but it deserves special attention as it certainly has a role, as shown
by \citet{2002MNRAS.336..785S}. By simply looking at the evolution of
these waves over time windows of $1$ to $2$~Gyr, we can nevertheless
qualitatively deduce the impact of the resonances that these waves
introduce into the disc. Several resonance species are at work in the
simulation.

Changes in the distribution function DF(\oJr,\oLz) are perfectly
identifiable. The two vertical tails, located at $\mathLz\approx 4000$
and $4500$, and $\mathJr \approx 100$, similar in shape to
\citet{2012ApJ...751...44S}, could be the signature of an ILR
scattering of each of the two waves present, the intermediate and the
external. However, as these tails appear progressively
(Fig.~\ref{fig:lz_jr}), we cannot exclude that the origin of this
scattering is a single wave which would reappear at a lower frequency
and/or greater radius, thus explaining the duplication. In addition, a
low-level inspection of DF(\oJr,\oLz) map shows other signatures of
the same type but at much lower levels, supporting the latter
hypothesis.  The depopulation of circular orbits ($\mathJr=0$),
accompanied by a redistribution in \Lz\ (cf. Sect.~\ref{sec:zeroevol}
and Fig.~\ref{fig:jr0}) is associated to this ILR scattering by
intermittent waves.  If the ILR of the intermediate spiral and/or the
outer waves are involved, the energy contained in the waves is
transferred to the particles. But, since it is not the entire disc
that is heated by this mechanism, it remains cold enough to allow the
regular resurgence of the mode(s).

With three sets of waves, we suspect that resonance overlaps in
physical space, such as \OLRb$-$\CRi\ and \UHRb$-$\ILRi\ or
$+1/4$~iS$-$\ILRs, may play an important role.  For a full efficiency
of this process, \citet{1988MNRAS.232..733S} have shown that two
patterns must overlapped over a radial range. But the theory does not
predict anything about the duration of the overlap(s). As far as we
know, there is nothing to prevent intermittent waves from interfering
non-linearly.  \citet{1988MNRAS.232..733S} suggested that the
corotation of the inner wave must coincide with the ILR of the outer
wave, which is the case \CRi$-$\ILRs\ at $t=3.16-4.24$~Gyr.  They
showed that this kind of resonance overlap would make the non-linear
interaction between the two patterns much more efficient. This is the
process advocated by \citet{2010ApJ...722..112M} for amplifying radial
migration.

It is questionable whether an $(m>0)-(m<0)$ overlap (e.g. intermediate
spiral outer $m\!=\!+4\,-$\ILRs, or any OLR$-$ILR overlap as for a
double-barred system in \citet{2015A&A...575A...7W}, would facilitate
the transfer of angular momentum to the outer regions. Indeed, it has
long been shown \citep{1972MNRAS.157....1L} that particles at any
$m>0$ resonance (as well as at the corotation) absorb \Lz\ (and $E$)
from the wave while those at any $m<0$ give \Lz\ (and $E$) to the
wave. When two resonances with opposite signs overlap, the angular
momentum and energy acquired by the particles at the $m>0$ resonance
of the inner high-$\Omega_\mathrm{p}$ wave can be transferred to the
outer low-$\Omega_\mathrm{p}$ wave through any $m<0$ resonance.

It remains to understand the wide vertical tail visible in
DF(\oJr,\oLz) around $\mathoLz=3000$, which is approximately the value
for the bar corotation at $t=10.54$~Gyr. The large values of
\Jr\ reached are the signature of the `hot' population. This
population should not see a significant variation of \Jr\ since it is
supposed to be scattered by the bar corotation. However, the
approximate calculation of \Jr\ (Eq.~\ref{eq:jrmean}) prevents a more
detailed analysis of this population. Indeed,
$\overline{\mathEr/\kappa}$ is very different from
$\overline{\mathEr}/\overline{\kappa}$ in this region. This is easily
explained because these particles explore a large fraction of the bar
and the disc, where $\kappa$ values are very different. The epicycle
approximation is therefore not valid here.

It would have been interesting to compare the results obtained with
real data, notably those concerning Milky-Way obtained by Gaia-RVS
\citep{2019MNRAS.484.3291T}, or recent models also expressed in the
action-angle domain \citep[e.g.][and reference
  therein]{2020ApJ...896...15F}. However, as mentioned several times,
\PFAA\ does not have the characteristics of Milky Way. Therefore, we
cannot yet explain the similarities or differences with these recent
semi-analytical modelling works. We postpone to a later article the
analysis of ongoing simulations actually dedicated to Milky Way.

\section{Conclusions}
\label{sec:conclusions}

Using the epicycle formulation of the radial action \Jr\ to calculate
the Chirikov diffusion rate and the associated characteristic
timescale, in addition to the results already obtained so far for
\Lz\ and $E$ by \citet{2020ApJ...889...81W}, we have shown that :

\begin{enumerate}
\item{The distribution of \Jr\ diffusion timescales is spiked around
  $T_\mathrm{D}(\mathJr)=0.9$~Gyr, mainly due to disc particles. It is
  followed by a plateau ending at $T_\mathrm{D}(\mathJr)\approx
  3$~Gyr those main contributor is the stellar bar. Roughly
  0.5~\mtot\ has $T_\mathrm{D}(\mathJr) < 3$~Gyr, 0.77~\mtot\ has
  $T_\mathrm{D}(\mathJr) < 10.54$~Gyr (i.e. the simulation length). Beyond
  the bar UHR, the space averaged timescale is $\langle
  T_\mathrm{D}(\mathJr)\rangle \sim 1$~Gyr.}
\item{By selecting particles as $T_\mathrm{D}(\mathJr) < 3$~Gyr and
  $T_\mathrm{D}(\mathLz) > 10.54$~Gyr, i.e. 0.25~\mtot, we identify
  all the particles that participate in the morphological structures
  characteristic of resonances and waves with the exception of the
  stellar bar.}
\item{Secular radial heating has been identified in the disc though
  the depopulation of circular orbits. 57\%\ of particles on circular
  orbits ($\mathJr=0$) at $t=3.16$~Gyr have \Jr\ increased by a few
  tens of \Lzunit\ after 7~Gyr (leading to $\sigma_\mathrm{R}$
  increase by $\sim 10$~\kms). This decircularisation is accompanied by
  \Lz\ transfer through coherent wave$-$particle interactions.}
\item{The signature of ILR scattering by disc wave(s) has been
  detected by ridges in the average distribution function
  DF(\oJr,\oLz), similarly to \citet{2012ApJ...751...44S}.}
\item{A wave analysis identified at least 3 types of waves: 1) a
  permanent stellar bar, 2) a non-permanent intermediate spiral which
  could be a beating phenomenon with 3) a set of intermittent
  multi-armed wave packets that carry \Lz\ towards the edge of the
  disc. Several resonance overlaps ensure a continuous cascade of waves
  which can also contribute to promote \Lz\ exchanges. But none of
  these overlaps is of the ILR$-$CR type.}  
\end{enumerate}

The Chirikov diffusion rate allows the separation of particle sets
with similar dynamic behaviour. It is a useful complementary tool for
dynamical analysis. Numerous avenues for further investigation will be
the subject of next articles: study of the bar, diffusion of $R$,
$\kappa$ and $X$, impact of live dark matter particles, etc. In
particular, we have calculated here $D_2$ based on an average of the
properties over 100~Myr, but other scales are possible
\citep{2020ApJ...889...81W}. It is through these other timescales that
possible effects due to chaos could appear
\citep{1992rcd..book.....L}.

\begin{acknowledgements}
I thank the anonymous referee for helpful comments. I would like to
acknowledge the Meso@LR computing centre of the University of
Montpellier for providing access to computing resources. This project
has been funded by a grant from the Scientific Council of the
University of Montpellier.
\end{acknowledgements}

\bibliographystyle{aa} 
\bibliography{diffusion_jr} 

\begin{thebibliography}{33}
\expandafter\ifx\csname natexlab\endcsname\relax\def\natexlab#1{#1}\fi

\bibitem[{{Binney} \& {Tremaine}(2008)}]{2008gady.book.....B}
{Binney}, J. \& {Tremaine}, S. 2008, {Galactic Dynamics: Second Edition}
  (Princeton University Press)

\bibitem[{{Buta}(2019)}]{2019MNRAS.488..590B}
{Buta}, R.~J. 2019, \mnras, 488, 590

\bibitem[{{Chirikov}(1979)}]{1979PhR....52..263C}
{Chirikov}, B.~V. 1979, \physrep, 52, 263

\bibitem[{{Frankel} {et~al.}(2020){Frankel}, {Sanders}, {Ting}, \&
  {Rix}}]{2020ApJ...896...15F}
{Frankel}, N., {Sanders}, J., {Ting}, Y.-S., \& {Rix}, H.-W. 2020, \apj, 896,
  15

\bibitem[{{Freeman}(1966)}]{1966MNRAS.133...47F}
{Freeman}, K.~C. 1966, \mnras, 133, 47

\bibitem[{{Halle} {et~al.}(2015){Halle}, {Di Matteo}, {Haywood}, \&
  {Combes}}]{2015A&A...578A..58H}
{Halle}, A., {Di Matteo}, P., {Haywood}, M., \& {Combes}, F. 2015, \aap, 578,
  A58

\bibitem[{{Kalnajs}(1971)}]{1971ApJ...166..275K}
{Kalnajs}, A.~J. 1971, \apj, 166, 275

\bibitem[{{Lichtenberg} \& {Lieberman}(1992)}]{1992rcd..book.....L}
{Lichtenberg}, A. \& {Lieberman}, M. 1992, {Regular and Chaotic Dynamics} (New
  York: Springer)

\bibitem[{{Lynden-Bell} \& {Kalnajs}(1972)}]{1972MNRAS.157....1L}
{Lynden-Bell}, D. \& {Kalnajs}, A.~J. 1972, \mnras, 157, 1

\bibitem[{{Mackereth} {et~al.}(2019){Mackereth}, {Bovy}, {Leung}, {Schiavon},
  {Trick}, {Chaplin}, {Cunha}, {Feuillet}, {Majewski}, {Martig}, {Miglio},
  {Nidever}, {Pinsonneault}, {Aguirre}, {Sobeck}, {Tayar}, \&
  {Zasowski}}]{2019MNRAS.489..176M}
{Mackereth}, J.~T., {Bovy}, J., {Leung}, H.~W., {et~al.} 2019, \mnras, 489, 176

\bibitem[{{Masset} \& {Tagger}(1997)}]{1997A&A...322..442M}
{Masset}, F. \& {Tagger}, M. 1997, \aap, 322, 442

\bibitem[{{Michel-Dansac} \& {Wozniak}(2006)}]{2006A&A...452...97M}
{Michel-Dansac}, L. \& {Wozniak}, H. 2006, \aap, 452, 97

\bibitem[{{Minchev} \& {Famaey}(2010)}]{2010ApJ...722..112M}
{Minchev}, I. \& {Famaey}, B. 2010, \apj, 722, 112

\bibitem[{{Minchev} {et~al.}(2011){Minchev}, {Famaey}, {Combes}, {Di Matteo},
  {Mouhcine}, \& {Wozniak}}]{2011A&A...527A.147M}
{Minchev}, I., {Famaey}, B., {Combes}, F., {et~al.} 2011, \aap, 527, A147

\bibitem[{{Patsis} {et~al.}(1994){Patsis}, {Hiotelis}, {Contopoulos}, \&
  {Grosbol}}]{1994A&A...286...46P}
{Patsis}, P.~A., {Hiotelis}, N., {Contopoulos}, G., \& {Grosbol}, P. 1994,
  \aap, 286, 46

\bibitem[{{Pfenniger}(1990)}]{1990A&A...230...55P}
{Pfenniger}, D. 1990, \aap, 230, 55

\bibitem[{{Pfenniger} \& {Friedli}(1991)}]{1991A&A...252...75P}
{Pfenniger}, D. \& {Friedli}, D. 1991, \aap, 252, 75

\bibitem[{{Ro{\v s}kar} {et~al.}(2012){Ro{\v s}kar}, {Debattista}, {Quinn}, \&
  {Wadsley}}]{2012MNRAS.426.2089R}
{Ro{\v s}kar}, R., {Debattista}, V.~P., {Quinn}, T.~R., \& {Wadsley}, J. 2012,
  \mnras, 426, 2089

\bibitem[{{Sanders} \& {Binney}(2014)}]{2014MNRAS.441.3284S}
{Sanders}, J.~L. \& {Binney}, J. 2014, \mnras, 441, 3284

\bibitem[{{Sanders} \& {Binney}(2016)}]{2016MNRAS.457.2107S}
{Sanders}, J.~L. \& {Binney}, J. 2016, \mnras, 457, 2107

\bibitem[{{Sch{\"o}nrich} \& {Binney}(2009)}]{2009MNRAS.396..203S}
{Sch{\"o}nrich}, R. \& {Binney}, J. 2009, \mnras, 396, 203

\bibitem[{{Sellwood}(2012)}]{2012ApJ...751...44S}
{Sellwood}, J.~A. 2012, \apj, 751, 44

\bibitem[{{Sellwood} \& {Binney}(2002)}]{2002MNRAS.336..785S}
{Sellwood}, J.~A. \& {Binney}, J.~J. 2002, \mnras, 336, 785

\bibitem[{{Skokos} {et~al.}(2002){Skokos}, {Patsis}, \&
  {Athanassoula}}]{2002MNRAS.333..847S}
{Skokos}, C., {Patsis}, P.~A., \& {Athanassoula}, E. 2002, \mnras, 333, 847

\bibitem[{{Soubiran} {et~al.}(2008){Soubiran}, {Bienaym{\'e}}, {Mishenina}, \&
  {Kovtyukh}}]{2008A&A...480...91S}
{Soubiran}, C., {Bienaym{\'e}}, O., {Mishenina}, T.~V., \& {Kovtyukh}, V.~V.
  2008, \aap, 480, 91

\bibitem[{{Sparke} \& {Sellwood}(1987)}]{1987MNRAS.225..653S}
{Sparke}, L.~S. \& {Sellwood}, J.~A. 1987, \mnras, 225, 653

\bibitem[{{Sygnet} {et~al.}(1988){Sygnet}, {Tagger}, {Athanassoula}, \&
  {Pellat}}]{1988MNRAS.232..733S}
{Sygnet}, J.~F., {Tagger}, M., {Athanassoula}, E., \& {Pellat}, R. 1988,
  \mnras, 232, 733

\bibitem[{{Trick} {et~al.}(2019){Trick}, {Coronado}, \&
  {Rix}}]{2019MNRAS.484.3291T}
{Trick}, W.~H., {Coronado}, J., \& {Rix}, H.-W. 2019, \mnras, 484, 3291

\bibitem[{{Vasiliev}(2019)}]{2019MNRAS.482.1525V}
{Vasiliev}, E. 2019, \mnras, 482, 1525

\bibitem[{{Wozniak}(2015)}]{2015A&A...575A...7W}
{Wozniak}, H. 2015, \aap, 575, A7

\bibitem[{{Wozniak}(2020)}]{2020ApJ...889...81W}
{Wozniak}, H. 2020, \apj, 889, 81

\bibitem[{{Wozniak} \& {Pfenniger}(1997)}]{1997A&A...317...14W}
{Wozniak}, H. \& {Pfenniger}, D. 1997, \aap, 317, 14

\bibitem[{{Zang} \& {Hohl}(1978)}]{1978ApJ...226..521Z}
{Zang}, T.~A. \& {Hohl}, F. 1978, \apj, 226, 521

\end{thebibliography}

\end{document}